\documentclass[%
 aip,
 amsmath,amssymb,
 reprint,%
]{revtex4-1}

\usepackage{graphicx}
\usepackage{placeins}
\usepackage{hyperref}
\usepackage{cleveref}
\usepackage{color,soul} 

\newcommand{\U}[2][\;]{\mathrm{#1 #2}}
\newcommand{\E}[1]{\cdot 10^{#1}}
\newcommand{\Eb}[1]{10^{#1}}

\newcommand{\dens}{m^{-3}}

\newcommand{\ebd}[1]{\mathrm{10^{#1}\; m^{-3}}}
\newcommand{\ed}[1]{\mathrm{\cdot 10^{#1}\; m^{-3}}}

\newcommand{\lb}{\left(}
\newcommand{\rb}{\right)}
\newcommand{\br}[1]{\lb #1 \rb} 

\newcommand{\lifu}{$\U{\mu V s^{-1}}$}

\newcommand{\pic}[4][1.0]{
\begin{figure}[htp]
	\begin{center}
		\includegraphics[width=#1\linewidth]{#2}
		\caption{#3}
		\label{#4}
	\end{center}	
\end{figure}}

\makeatletter
\def\@email#1#2{%
 \endgroup
 \patchcmd{\titleblock@produce}
  {\frontmatter@RRAPformat}
  {\frontmatter@RRAPformat{\produce@RRAP{*#1\href{mailto:#2}{#2}}}\frontmatter@RRAPformat}
  {}{}
}%
\makeatother

\begin{document}

\title{Exploration of Helicon Plasmas for Wakefield Accelerators at the Madison AWAKE Prototype}

\author{Marcel Granetzny}
\email{granetzny@wisc.edu}
\author{Barret Elward}
\author{Michael Zepp}
\author{Maxwell Loughan}
\author{Oliver Schmitz}

\affiliation{Department of Nuclear Engineering and Engineering Physics, University of Wisconsin - Madison, Madison, Wisconsin, 53706, USA}

\date{\today}

\begin{abstract}
Plasma wakefield accelerators have the potential to revolutionize particle physics by providing lepton collision energies orders of magnitude beyond current technology. Crucially, these accelerators require a high-density, highly homogeneous, scalable plasma source. The Madison AWAKE Prototype (MAP) is a new plasma development platform that has been built as part of CERN's beam-driven wakefield accelerator project AWAKE. MAP uses a dual helicon antenna setup with up to 20 kW of RF power to create plasmas in the low $\ebd{20}$ range in a highly uniform magnetic field. The project is supported by a range of diagnostics that allow non-invasive measurements of plasma density, ion and neutral flows, and temperatures, and a 3D finite element model that can calculate helicon wavefield and power deposition patterns. In this paper, we present an in-depth overview of MAP's design and construction principles and main physics results. We show that the plasma discharge direction is set by the combination of antenna helicity and field direction and linked to the well-known preference for right-handed helicon modes. We find that the plasma density depends dramatically on the direction of plasma and neutral flow. A detailed measurement of the ionization source rate distribution reveals that most of the plasma is fueled radially by recycling at the wall, a finding with strong implications for optimizing plasma homogeneity. Lastly, we describe how helicon antennas can be engineered to optimize power coupling for a given target density. Together these findings pave the way toward the practical use of helicon plasmas in wakefield accelerators.
\end{abstract}


\maketitle

\section{Introduction}

Plasma wakefield accelerators could achieve acceleration fields orders of magnitude higher than current state-of-the-art technology and enable compact lepton colliders with collision energies in the TeV range. One of the core challenges in developing this technology is engineering a high-density, highly uniform, scalable, and reliable plasma source. In this paper, we describe the design, construction, and current results of the \textit{Madison AWAKE Prototype} (MAP), a dedicated plasma development platform based on helicon waves.\\

In the rest of this introduction, we will briefly review plasma wakefield acceleration and introduce the AWAKE project and MAP. \Cref{sec:MAP} features an in-depth description of MAP's design and construction principles as well as an overview of MAP's diagnostics. The goal is to provide a blueprint for the design of other high-power helicon devices. \Cref{sec:firstResultsMAP} is an overview of experiments and studies performed on MAP. This includes characterization of initial and high-power plasmas, operation with multiple helicon antennas, plasma directionality control, particle and power balance, density and power coupling optimization, and more. We conclude with a summary and descriptions of planned upgrades and experiments in \cref{sec:summary}.

\subsection{Plasma Wakefield Acceleration}
Current-day particle accelerators at the highest energy frontier are proton colliders. However, the composite nature of protons massively complicates data analysis and interpretation. Fortunately, such difficulties are of no concern when colliding fundamental particles such as leptons. TeV lepton accelerators would therefore be highly promising candidates for next-generation collider experiments and have sparked major interest in the international particle physics community.\\

Muon colliders have been proposed\cite{accettura2023towards}, but suffer from the unstable nature of muons, which necessitates producing them artificially in large quantities. An electron collider seems the most practical, but electron energies in circular colliders are severely limited by synchrotron radiation. For example, the Large Hadron Collider (LHC) has a circumference of 27 km and can accelerate protons up to 7 TeV, but synchrotron losses would limit electrons to much lower energies. Prior to the LHC's construction, its tunnel was used to house the Large Electron Positron Collider (LEP), which achieved maximum collision energies of 209 GeV.\cite{assmann2001lep}\\

A future TeV lepton collider will therefore need to be linear instead of circular. Unfortunately, while a circular collider can accelerate particles during an unlimited number of passes, a linear collider needs to supply the final collision energy during a single pass. Currently used superconducting radio frequency cavities can provide fields up to $100\U{MV\, m^{-1}}$.\cite{Caldwell2016} A future lepton collider based on this technology would therefore have to be around $100 \U{km}$ long to achieve a center of mass energy of 10 TeV. Such a large experiment is likely to feature massive costs and long construction times, potentially making it unfeasible altogether. However, a first-generation plasma wakefield accelerator (PWA) could provide gains over $1\U{GeV\, m^{-1}}$. PWAs would therefore allow for dramatic cost reductions with the prospect of ultimately increasing collision energies by several orders of magnitude.\cite{tajima2020wakefield,gschwendtner2019plasma}.\\

PWAs use a linear high-density plasma with very high on-axis homogeneity to accelerate particles. This plasma is then perturbed by a driver, such as an energetic proton or laser beam, that induces a relativistic Langmuir wave in its wake. Since Langmuir waves induce a large charge imbalance in the plasma, they create strong electric fields in the driver's wake. These wakefields can be used to accelerate so-called witness particles such as electrons or positrons. The maximum achievable field is the wave breaking field $E_{wb}$ and depends directly on the plasma density $n_e$ as \cite{Dawson_1959}

\begin{align}
\label{eq:WBField}
E_{WB} &= \frac{\omega_{pe} m_e c}{e} \approx \sqrt{\frac{n_e}{\ebd{20}}}\U{GV m^{-1}}.
\end{align}

PWAs can be divided into two main types: laser-driven and beam-driven. Laser-driven PWAs have demonstrated accelerating fields of $50\U{GV\, m^{-1}}$\cite{blumenfeld2007energy}. However, these fields have only been achieved on sub-meter length scales and it is unclear whether this approach can be scaled up to build a TeV lepton collider at sufficiently high luminosity\cite{herr2006concept} for practical particle physics studies. Beam-driven PWAs provide an alternative that has the potential to scale better, albeit at lower achievable acceleration fields in the low $\U{GV\, m^{-1}}$ range. Importantly, PWAs of either type require a source of highly ionized plasma since any remaining neutral atoms will lead to energy loss and scattering of the particles to be accelerated. In addition, the plasma core needs to be highly homogeneous to keep the wakefields in phase with the witness particles.\\

\subsection{The AWAKE Project and MAP}

The \textit{Advanced Proton Driven Plasma Wakefield Acceleration Experiment} (AWAKE) at CERN\cite{Caldwell2016,gschwendtner2016awake} has demonstrated acceleration of electrons in $\U{GV\, m^{-1}}$ fields over a range of ten meters. The goal is to produce electron bunches with TeV energies in a future full-scale accelerator. However, the plasma source currently in use is a 10-meter-long, laser-ionized rubidium vapor cell. Due to laser attenuation, this approach is not transferable to a full-scale facility in the hundreds of meter range and beyond. Therefore, AWAKE requires a new plasma source that is scalable and can reliably create highly uniform plasmas in the $\ebd{20}$ range.\cite{Muggli_2018}\\

One of the most promising candidates is plasma breakdown and sustainment by helicon waves\cite{Boswell1984,Chen2015}. Helicons are magnetized plasma waves in the radio frequency range, typically at tens of MHz. Densities in the $\Eb{20}\U{\dens}$ range have been achieved in a $1$ m long, high-power helicon plasma\cite{stollberg2024first,Buttenschon2018}. Longer helicon devices have been constructed, for example, MARIA\cite{Green2020}, PHASMA\cite{Shi_2021} and RAID\cite{Furno_2017}, but these devices do not reach sufficiently high densities for accelerator applications. In addition, helicon plasmas usually have significant axial density variations.\cite{Chen2015}\\

In a helicon plasma, the wave fields create localized heating patterns and result in a self-consistent equilibrium between power deposition, ionization, and various particle and energy flows and loss channels. After a few milliseconds, this process leads to stable density, temperature, and neutral profiles. By first understanding and then shaping the RF power deposition in a helicon plasma this process can be influenced to optimize a plasma for high density and axial homogeneity. The \textit{Madison AWAKE Prototype} (MAP) has been built to optimize an argon plasma for use in AWAKE. MAP is unique in its combination of axial field uniformity, high-power operation with multiple antennas, diagnostic capabilities, degree of automation, and overall flexibility, which make it an ideal platform for accelerator-relevant and fundamental helicon studies.

\section{Experiment}
\label{sec:MAP}

This section contains a detailed description of the Madison AWAKE Prototype (MAP) and diagnostics. We describe the mechanical design and magnetic field setup in \cref{sec:overview,sec:magSetup}. \Cref{sec:RFChain,sec:antennas,ssec:faraday} highlight the RF setup, antenna design, and Faraday screen. The current suite of diagnostics and control systems, including interferometry and laser-induced fluorescence is described in \crefrange{sec:interferometer}{sec:controlInterface}.

\subsection{Overview}
\label{sec:overview}

\begin{figure*}
    \centering
    \includegraphics[width=1.0\linewidth]{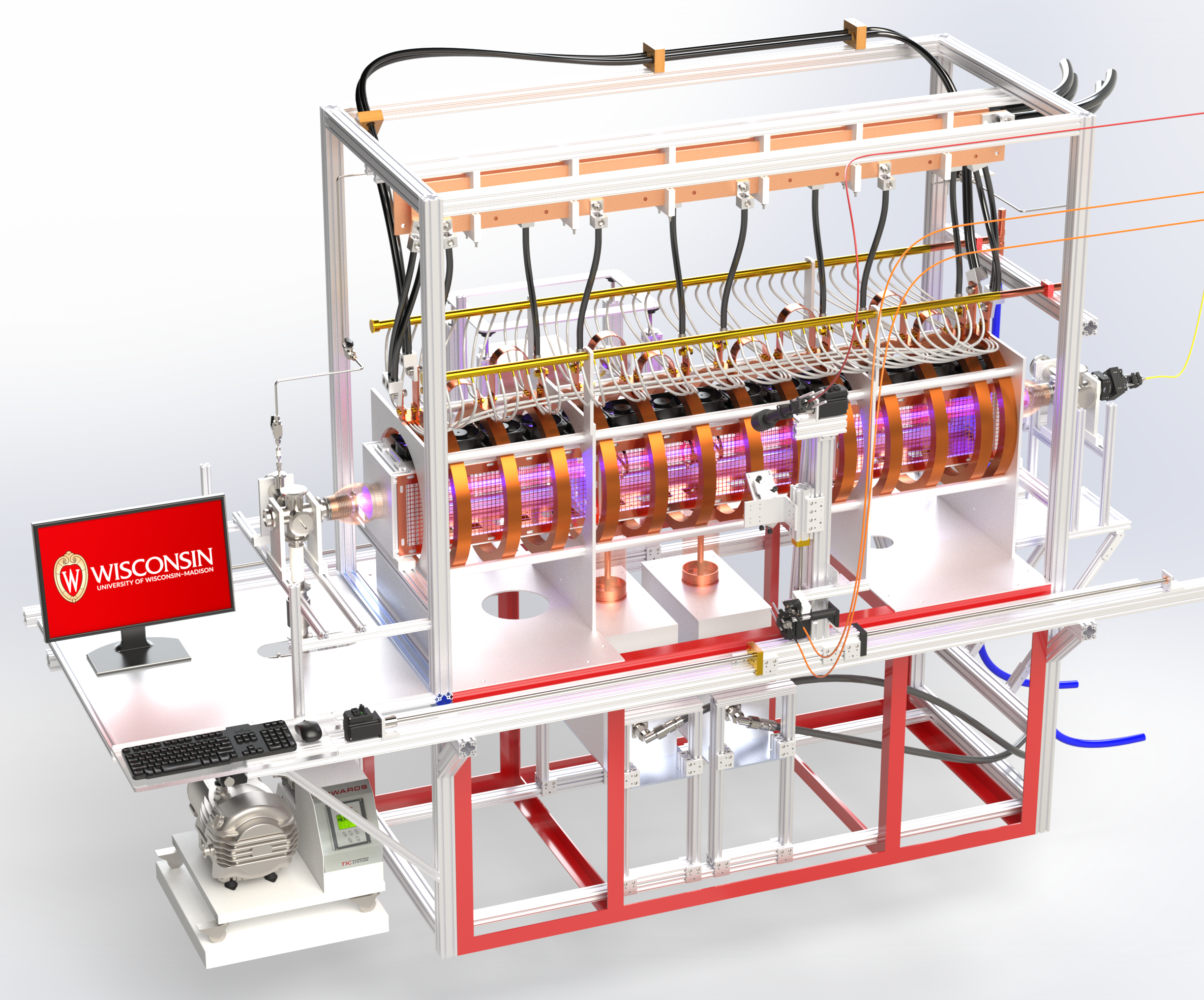}
    \caption{CAD model of the Madison AWAKE Prototype. 14 magnets create a uniform 50 mT field inside the borosilicate vacuum vessel. The argon plasma is fueled from the left or right side and sustained by two helicon antennas that can be positioned at 11 different axial positions. The plasma is surrounded by a gapless Faraday screen inside the magnets. Laser-induced fluorescence, heterodyne microwave interferometry, and passive spectroscopy are mounted on motorized 2-axis motion platforms. The experiment uses a modular control interface that allows synchronized and scripted firing, data acquisition, and diagnostic positioning.}
    \label{fig:MAPCAD}
\end{figure*}

 A CAD model of MAP is shown in \cref{fig:MAPCAD}. The vacuum vessel is made from a $2.6$ m long 1borosilicate glass tube with an inner diameter of $52$ mm and an outer diameter of $56$ mm. Each end terminates in a 6-way stainless steel KF cross, allowing for pumping, gas inlet, and diagnostic access. Argon gas can be fed in on either or both ends. The main part of the vacuum vessel rests inside a set of $14$ water-cooled copper coils that can produce fields up to 50 mT.\\
 
 A Faraday screen consisting of detachable panels surrounds the vacuum vessel inside the magnetic field coils. The vacuum vessel is air-cooled by an array of fans on the top side of the Faraday cage to allow for continuous, hours-long plasma operation. The Faraday screen has flexible feedthroughs at the bottom to allow the use of multiple helicon antennas at 11 different axial positions. The detachable screen panels allow for quick adjustments to the antenna setup such as changing the helicity or antenna type. Power to these antennas is fed from RF generators and passed through impedance-matching networks. The antennas themselves are of the half-helical type\cite{Miljak1998}. As an example, \cref{fig:MAPCAD} shows a setup with two antennas spaced 30 cm apart close to the middle of the magnetic field.\\
 
   The main diagnostics on MAP are laser-induced fluorescence, heterodyne microwave interferometry, and passive spectroscopy. These diagnostics are mounted on motor-driven carts that enable computer-controlled positioning. The RF generators and diagnostics are computer-controlled as well, which allows for automated discharge and diagnostic triggering. In conjunction with the motion control system, this capability enables us to perform automated measurement campaigns that can diagnose the entire MAP plasma. The following sections describe the different subsystems in detail.\\

\subsection{Magnetic Field Setup}
\label{sec:magSetup}

\begin{figure}[htp]
			\includegraphics[width=\linewidth]{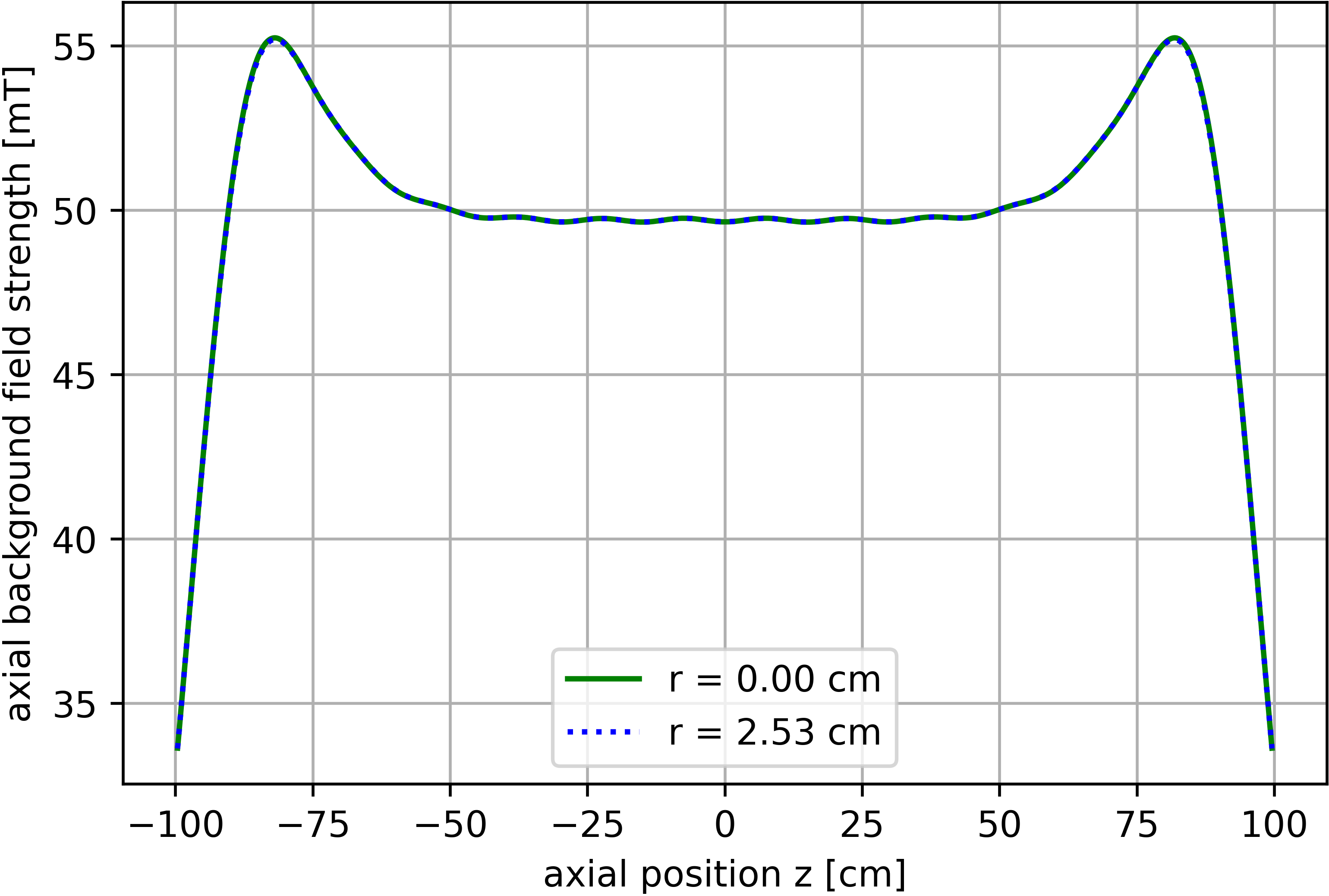}
		\caption{Magnitude of the magnetic field in MAP for 222 A through the inner 12 coils and 333 A through the 2 end coils calculated with a COMSOL finite element model. The field is identical at the core and edge of the vacuum vessel and highly homogenous in the central section.}
	\label{fig:comMag}	
\end{figure}

The coils shown in \cref{fig:MAPCAD} have an inner diameter of $40$ cm, an outer diameter of $47$ cm, consist of $27$ turns each, and use in-conduit water cooling. The inner $12$ coils are spaced 14.9 cm apart while the two end coils are spaced 8.1 cm away from the last inner coils. To achieve a uniform field,  the central $12$ coils are divided into $4$ coil groups, each consisting of $3$ coils connected in series. These $4$ groups are connected in parallel to the bus bars at the top of MAP. The two end coils are connected in series with each other, and in parallel with the other four coil groups to the bus bars. The bus bars are fed from a single DC power supply that outputs up to 1220 A. This allows for an inner coil current of 222 A and an end coil current of $333$ A. As shown in \cref{fig:comMag}, this setup enables a constant $50$ mT field in the central 1 m section of MAP and produces a 1.8 m wide region with a nearly homogeneous field of $50 \pm 5 \U{mT}$. As shown in \cref{fig:comMag}, radial variations in the field strength throughout the vacuum vessel are negligible. The wiring setup, bus bars, and cooling lines are shown in \cref{fig:MAPCAD}.

In order to allow for easy changing of the magnetic field direction we use the jumper system shown in \cref{fig:magBox}. Currents from the magnet power supply enter from the top right and return to the bottom right. The magnet bus bars are fed through conductors entering the box on the left. The direction of the magnetic field can be chosen by connecting the power supply side posts to the magnet side posts through the set of two large jumper bars. In \cref{fig:magBox} the upper and lower jumper bars are crossed, resulting in a field pointing to the left in \cref{fig:MAPCAD}. The jumpers can be arranged parallel to each other to reverse the magnetic field. The lower jumper bar is accessed through a small panel at the bottom of the box. This simple and compact setup allows studies under regular and reverse field conditions at 1220 A of total current with changeover times of only a few minutes.\\

\pic{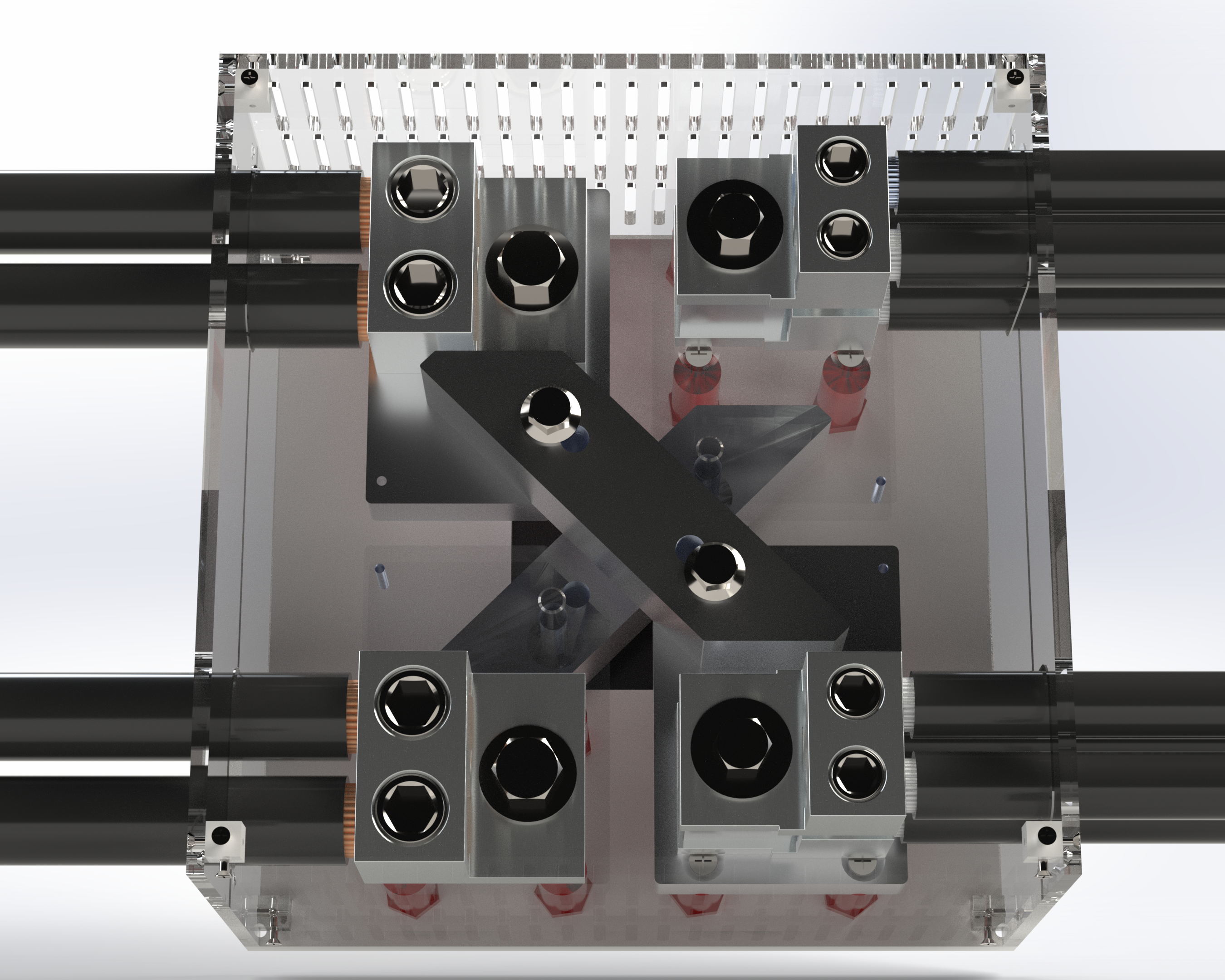}{CAD model of the jumper box used to set the magnetic field direction. Aluminum jumpers on the top and bottom are used to direct current entering from the power supply on the right towards the magnets connected on the left. The magnetic field direction can be set by either crossing the jumpers as shown or arranging them in parallel. The lower jumper is accessed through the black bottom panel.}{fig:magBox}

\subsection{RF Drive Chain}
\label{sec:RFChain}

The helicon antennas on MAP are driven from water-cooled, 10 kW RF power supplies running at 13.56 MHz. Using RG218/U coaxial cable, RF power is passed from the generators to impedance-matching networks directly under the antennas. Each matching network employs two high-voltage capacitors and an inductor in an L-type configuration. The capacitors can be adjusted through electrical motors inside the network, which allows for remote-controlled plasma impedance matching during live power operation directly from the control station.\\

From the matching network, a PTFE and copper coaxial transmission line covers the final leg to the helicon antenna. This design was chosen to minimize radiated RF emissions originating between the matching network and antenna. PTFE is used as an insulator to prevent melting due to high reflected power during the matching process, which was found to occur at the MARIA\cite{Green2020} experiment when using a nylon-based transmission line. The coaxial design further prevents arcing to the Faraday screen, which was a major issue in previous designs involving strip lines. Lastly, a tubular copper mesh is used to connect the bottom of the Faraday screen power feedthrough, see \cref{ssec:faraday}, to the top of the matching network to capture any RF emissions from this opening.\\

The RF drive chain close to the plasma is shown in detail in \cref{fig:MAPRFChain}. RF power is fed into the matching networks from the lower right and the coaxial transmission lines connect the matching networks to the antennas after passing through the Faraday screen.\\

\pic{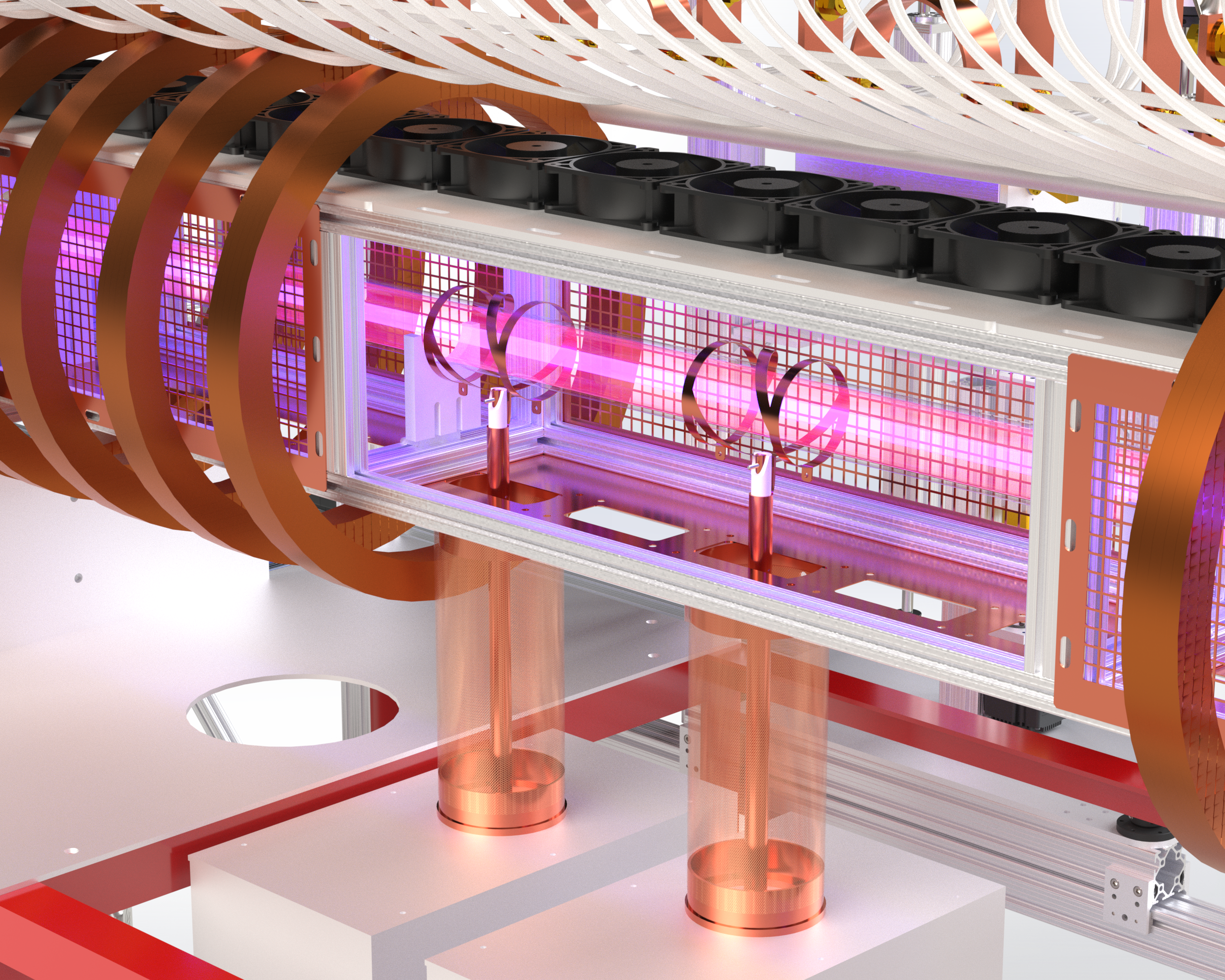}{Close-up of the Faraday screen and RF chain for a dual antenna setup, with some magnets and structural supports hidden for better visibility. RF power is fed through RG218/U cables into the bottom left side of each matching network, as shown in \cref{fig:MAPCAD}. A rigid copper and PTFE coaxial transmission line passes from each matching network up through openings in the Faraday screen and supports the half-helical antennas. A tubular copper mesh between the Faraday screen and matching networks catches any RF emissions leaking through the screen feedthrough. As shown here, the Faraday screen panels can be removed individually to allow access to the antennas. The bottom of the screen has multiple openings for antenna transmission lines. These feed-throughs are covered when not in use but shown open here for better visibility. Antennas have a 1 cm air gap to the vacuum vessel to prevent excessive heating while the fan array on the top provides cooling to the vacuum vessel.}{fig:MAPRFChain}

Two sets of power supplies, matching networks, and antennas are currently in use and additional sets might be added in the future. The generators are computer-controlled, which allows us to set individual power levels or pulse profiles and enables fire synchronization. The setup further allows for arbitrary phasing between the different RF power supplies through an external reference function generator.\\

\subsection{Antenna Design}
\label{sec:antennas}
The antennas currently in use are 10 cm long and use a half-helical design\cite{Miljak1998}, which is known to excite the highest plasma densities\cite{Sudit1996}. Helicon antennas are commonly soldered together from individual strips of copper, a time-consuming process that makes the antennas very stiff. Changing such a stiff antenna then necessitates breaking the vacuum to slide a new antenna over the end of the tube or soldering the antenna together directly on the vacuum vessel. In contrast, MAP's antennas are cut from a sheet of $0.76\U{mm}$ ($0.03\U{in}$) thick copper using a water jet. Such an antenna cutout is demonstrated on the left in \cref{fig:flexAntenna}. The antenna is very flexible and can therefore easily be bent around the vacuum vessel by hand. This also allows for bending the same antenna into either a left- or right-handed helicity, the latter of which is demonstrated on the right in \cref{fig:flexAntenna}. At the same time, the antenna is still rigid enough to prevent deformation when heating up during steady-state operation over multiple hours.\\

The azimuthally open ends of the transverse straps are bolted together and the central taps on the helical straps connect directly to the transmission line leading to the matching network, as shown previously in \cref{fig:MAPRFChain}. This flexible design allows for quick changes to the antenna's helicity or size. During steady-state operation, the antenna and consequently vacuum vessel can reach significant temperatures that can overheat the vacuum vessel. To alleviate this problem we maintain a 1 cm radial air gap between the antenna and vacuum vessel as demonstrated in \cref{fig:MAPRFChain}. The chosen antenna thickness ensures that the antennas keep their shape and maintain the air gap even when heating up during steady-state operation over multiple hours. For example, the antenna on the right on \cref{fig:flexAntenna} was used for the majority of the studies in this paper and maintained its shape throughout as shown.\\

\pic{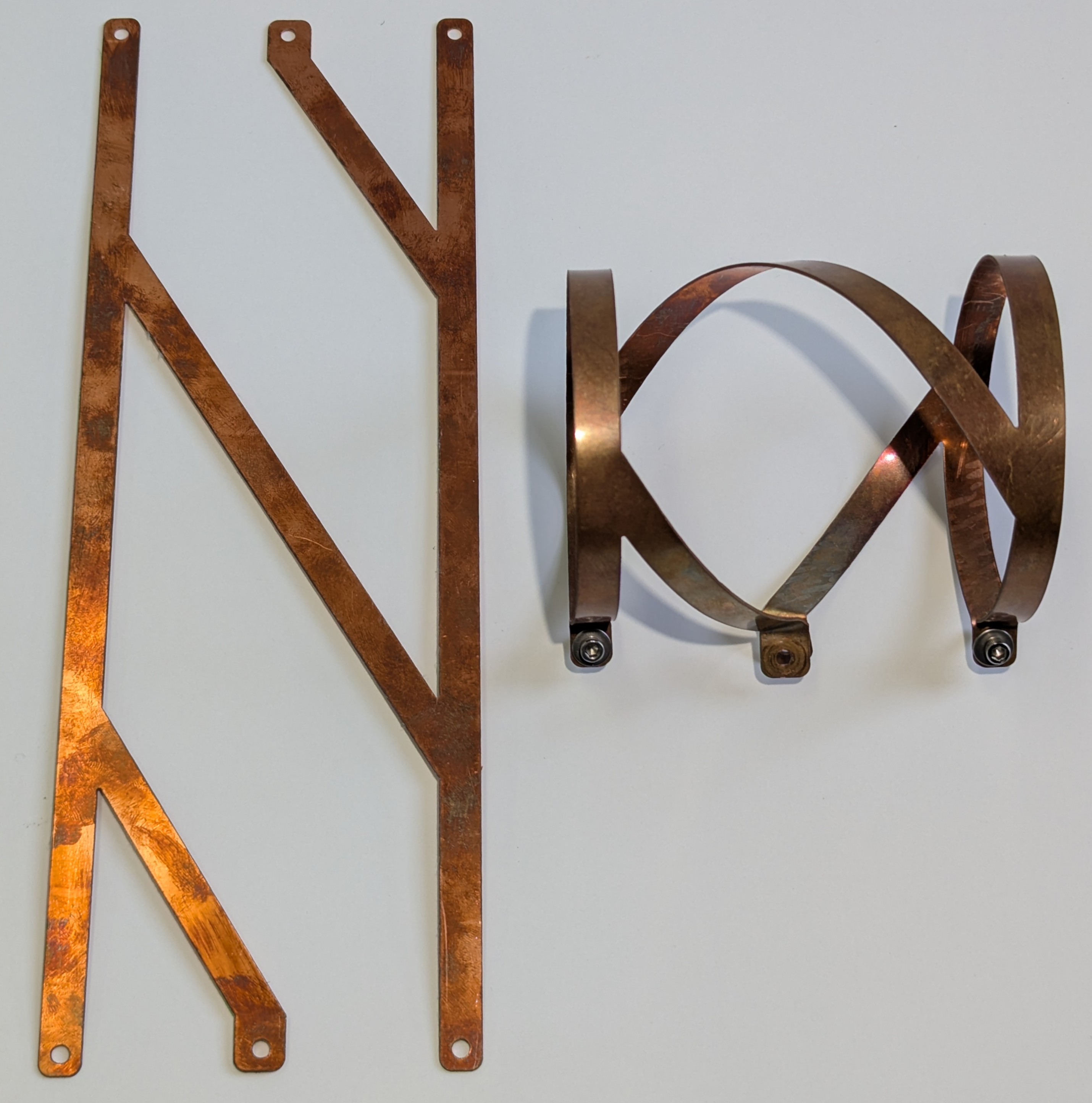}{Example of MAP antennas before and after bending. The antenna on the right has been used in the majority of studies in this paper. The end hoops are closed using fasteners and the middle tabs connect directly to the transmission line as shown in \cref{fig:MAPRFChain}. We use steel fasteners to allow for greater clamping force and ensure a reliable RF connection between the copper parts.}{fig:flexAntenna}

\subsection{Faraday Screen}
\label{ssec:faraday}

We noticed during early MAP operation that significant amounts of RF power was radiated into the rest of the laboratory even for a well-matched plasma. This led to significant noise problems for diagnostics and plasma controls. To solve this issue we constructed the Faraday screen shown in \cref{fig:MAPCAD,fig:MAPRFChain}. The screen is supported by an aluminum frame that rests inside the magnetic field coils and supports 12 detachable panels.\\

The three bottom panels are solid copper and provide a total of 11 openings for RF power feed-throughs. Each opening has a separate cover to eliminate RF leakage from any openings not in use. As mentioned in \cref{sec:antennas}, any openings in use are connected by tubular copper mesh to the matching network below to capture any remaining RF emissions. The top of the Faraday screen frame is covered by a continuous copper mesh under compression by three aluminum panels that contain five fans each. These fans blow air through the mesh to cool the antennas and the vacuum vessel during steady-state operation. Six panels make up the sides of the Faraday screen and feature a wider mesh with 12.7 mm spacing to allow measurement access for the microwave interferometer and LIF systems described later in \cref{sec:interferometer,sec:LIF}.\\

Supports for the vacuum vessel are mounted directly onto the Faraday screen frame and are made from glass-filled PTFE. At the axial ends, the Faraday screen consists of aluminum endplates that reduce the square frame cross-section from a side length of 254 mm to a circular cross-section with a diameter of 89 mm. From there a tubular copper mesh is used to electrically connect the screen to vacuum bellows at the ends of the borosilicate vessel. The screen is connected to MAP's base plate at its left and right ends using wide grounding straps.\\

This design ensures that the entire Faraday screen is free of any cuts, slits, or openings. At the same time, the detachable panels allow for quick modifications of the antenna setup. Together with the flexible antenna design described in \cref{sec:antennas}, this allows for a complete change of antenna helicity, length, type, or position in about 10 minutes.\\

To test the efficiency of the Faraday screen, we took RF field strength measurements using a close field antenna probe at the six locations indicated in the top panel of \cref{fig:noiseMeas} before and after installing the screen. These measurements were taken approximately at the height of the plasma column. We held the probe in air at locations 1, 2, and 3. At location 4 we placed the probe on a support beam near a Faraday screen grounding strap. At location 5 we held the probe directly against the screen and at position 6 we placed it on the aluminum baseplate. This test was performed at 500 W RF power using a 19 cm long antenna with the background field turned off, resulting in an axially symmetric capacitively coupled plasma.\\

\pic{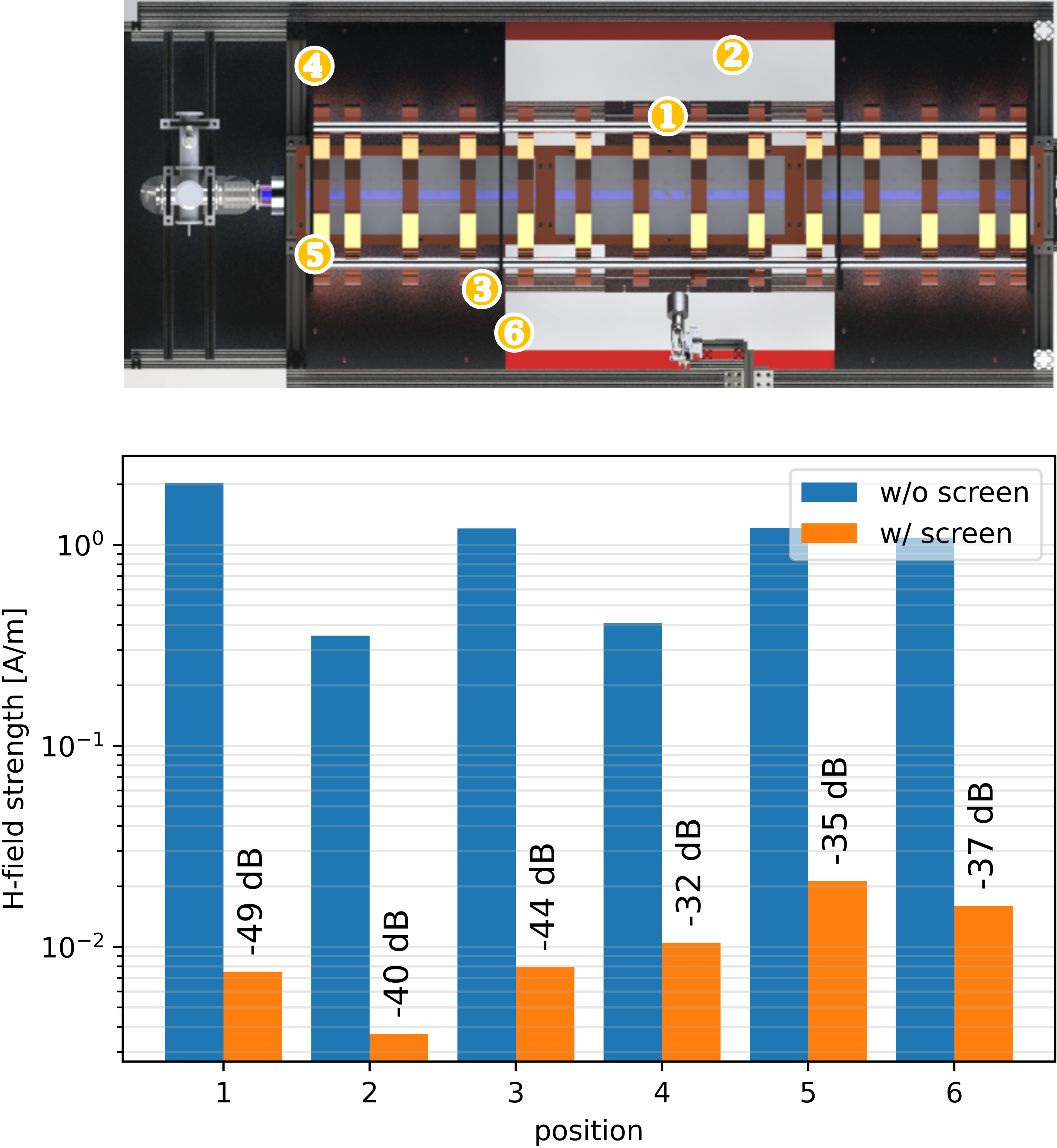}{Faraday screen effectiveness measurements. Top: Measurement locations on a simplified top view of MAP. Locations 1, 2, and 3 are in the air. Locations 4, 5, and 6 are near the left Faraday screen grounding strap, directly at the screen and on the grounded base plate, respectively. Bottom: Measured RF emissions before and after Faraday screen installation on a logarithmic scale with relative reductions in decibels. In the air, RF emissions amplitudes are reduced up to 270-fold.}{fig:noiseMeas}
 
 The test results are shown in the bottom panel of \cref{fig:noiseMeas} on a logarithmic scale. The Faraday screen reduces emissions in the air between 42 to 49 dB. Directly at the left grounding strap, the screen, and the baseplate, emissions are reduced by 32 to 37 dB. Together these measurements show that the Faraday screen succeeds in capturing RF emissions from the antenna and directing them from the screen to the ground plates and from there back to the RF power supply. The screen reduces RF field amplitudes in air up to 270-fold.

\subsection{Microwave Interferometry}
\label{sec:interferometer}

\begin{figure*}
    \centering
    \includegraphics[width=1.0\linewidth]{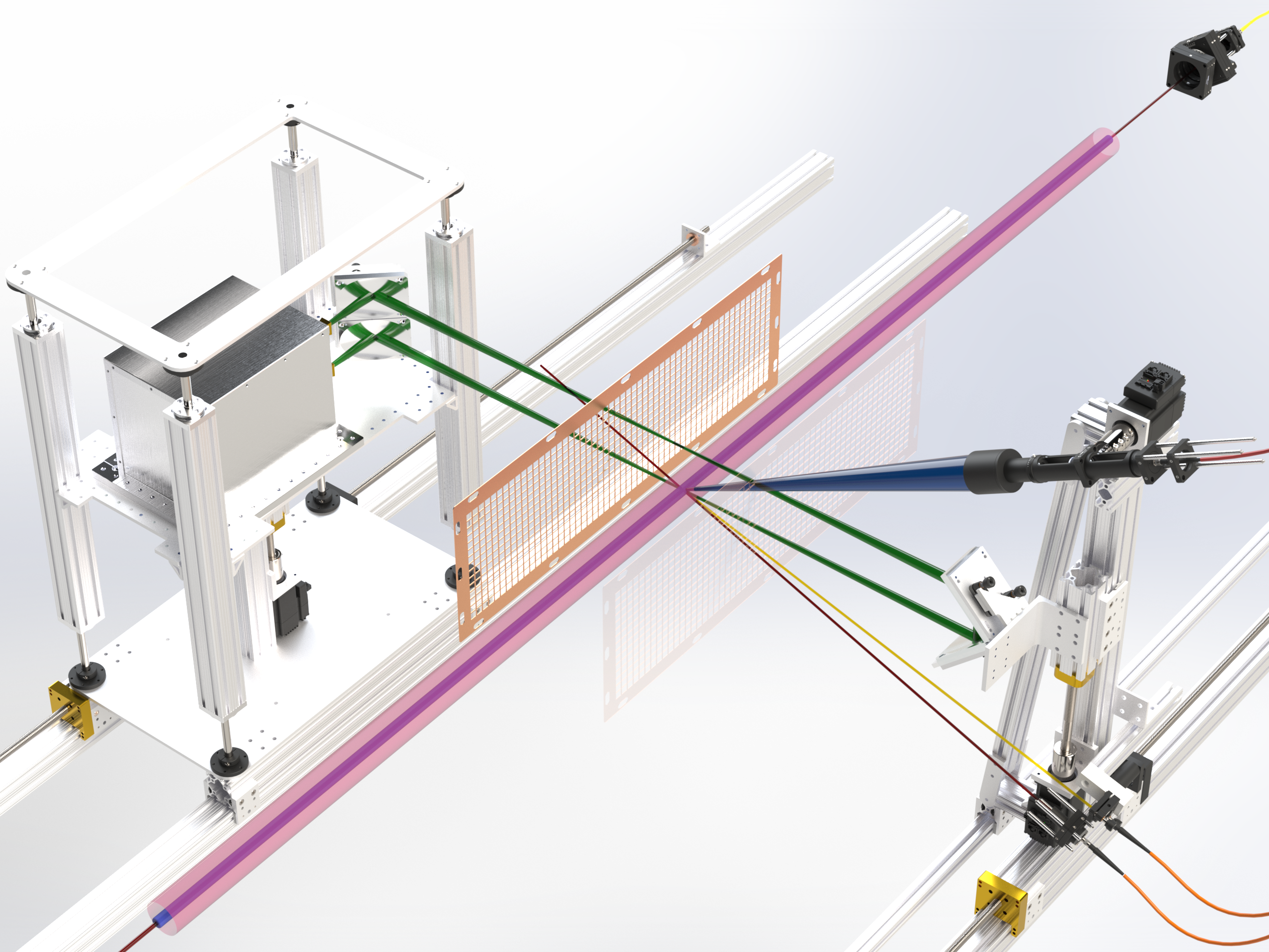}
    \caption{The MAP diagnostics and their beam paths through the plasma and Faraday screen. The interferometer is mounted on the 2-axis motion platform on the left. The 105 GHz beam, visualized in green, is focused by the lower elliptical mirror on the left to a point in the center of the plasma. A planar and elliptical mirror combination on the right sends the beam back to the interferometer enclosure 76 mm above the plasma center and through the Faraday screen. Lastly, the upper elliptical mirror on the left directs the beam into the receiving horn antenna. The laser-induced fluorescence system uses either axial or radial injection, both of which are shown by red beams originating from the top right and lower right, respectively. The fluorescence light, visualized in blue, is captured by the telescope on the middle right. Passive spectroscopy captures plasma emissions along the yellow line-of-sight. The diagnostics platforms can be moved axially and vertically through a system of four computer-controlled motors driving acme screws. The telescope is rotated using a fifth motor to sample different radial positions. This setup allows measurements at any axial and radial location in MAP, except those directly under a field magnet.}
    \label{fig:diagOverview}
\end{figure*}

MAP uses a heterodyne interferometer\cite{hartfuss1997heterodyne} system based on a 105 GHz microwave source with a cut-off density of $1.4\E{20}\U{\dens}$. The interferometer beam path through the MAP plasma is shown in green in \cref{fig:diagOverview}. The lower elliptical mirror on the left sends the beam towards the center of the plasma. A planar and elliptical mirror combination on the right directs the beam back to the interferometer enclosure 76 mm above the plasma center. The mirror system keeps the beam focused throughout and minimizes the beam waist inside the plasma. At 2.9 mm, the beam has a wavelength that experiences only minimal attenuation during the four passes through the Faraday screen side panels, which have a mesh size of 12.7 mm.\\

Measurement data is sent directly from the interferometer enclosure to the MAP control computer. The enclosure contains a complex combination of microwave components, FPGA logic circuitry, and a high-speed high-precision mixed signal circuit. Together with the quasi-optical mirror system, this setup achieves a time resolution of $5 \U{\mu s}$ during non-stop data streaming and a line-averaged density resolution of $1.5\E{17}\U{\dens}$ without the use of a digitizer, while also avoiding frequency drift issues.\cite{hartfuss1997heterodyne} This system will be described in more detail elsewhere.\cite{granetzny2025interferometer}\\

The interferometer enclosure is mounted on a 2-axis motion platform that is motor-driven and computer-controlled. A similar but smaller system is moving the return mirror system. This setup allows the positioning of the interferometer to measure the line-integrated density at any axial or vertical position in the MAP plasma, except those directly in the shadow of a magnet.\\

As shown later, for example in \cref{fig:powerScan}, MAP reaches densities up into the low $\ebd{20}$ range, which results in significant beam refraction. The main application of the interferometer is therefore to provide an absolute density calibration for the LIF system in the low to mid $\ebd{19}$ range. The LIF system is then used to measure radial density profiles up into the low $\ebd{20}$ range. We plan to upgrade the current interferometer setup with a terahertz source to allow direct density measurements at AWAKE target densities in the mid $\ebd{20}$ to low $\ebd{21}$ range.

\subsection{Laser Induced Fluorescence}
\label{sec:LIF}

MAP uses a laser-induced fluorescence (LIF) diagnostic nearly identical to the system used on MARIA\cite{Green_2019}\cite{Green_Schmitz_Zepp_2020}, with only minor modifications to increase the signal-to-noise ratio. The system measures the ion and neutral velocity distribution functions (VDF). Analysis of the VDF allows us to calculate ion and neutral temperatures and flow velocities. We are further able to measure the plasma density by calibrating the LIF system against other plasma diagnostics, such as the microwave interferometer.\\

The diagnostic is designed around a 40 mW tunable diode laser, amplified as high as 500 mW. The laser can excite the argon ion transition at 668.614 nm from the $3$d$^4$F$_{7/2}$ state to the $4$p$^4$D$_{5/2}$ state, which then decays with emission at 442.6 nm to the $4$s$^4$P$_{3/2}$ state within nanoseconds. The laser can also be tuned to excite the neutral argon transition at 667.9 nm, from the 4s$^2$[3/2]$_1$ state to the 4p$^2$[1/2]$_0$ state, which decays quickly with emission at 750.6 nm to the 4s$^2$[1/2]$_1$ state.\\

The laser is injected either axially or radially, as shown by the red beams on the top and bottom right in \cref{fig:diagOverview}. For radial injection, the laser is linearly polarized. For axial injection, a quarter waveplate and linear polarizer are used, so that the beam can be linear, right-hand circular, or left-hand circular polarized. Light is collected from the plasma through collection optics that look into the plasma radially, as visualized by the blue beam \cref{fig:diagOverview}. The collection optics are on a motorized fine rotation that enables radial scans in increments as small as 0.04 mm. The collected light is sent through an optical fiber to a photomultiplier tube (PMT). Subsequently, the fluorescence signal is extracted with a lock-in amplifier.\\

To measure the ion or neutral VDF, the laser frequency is scanned over a range on the order of 10 GHz. The amplitude of the ion VDF as output by the lock-in amplifier can be integrated to yield the LIF intensity $I$, in \lifu. $I$ is related to the plasma density $n$ through a scaling law of the form $n=AI^B$, where $A$ and $B$ are empirically derived constants. To calculate $B$, we used an RF-compensated Langmuir probe on the MARIA device over a wide range of plasma density and magnetic field values as described in \cite{Green_Schmitz_Zepp_2020}. $A$ is derived by integrating radially resolved LIF measurements on MAP and comparing the result to line-integrated interferometer measurements at the same location. Using this technique, we found $A = \br{1.85 \pm 0.27}\ed{18}$ and $B = 0.30 \pm 0.02$.

The velocity distribution function is also used to resolve ion and neutral flows. Flow velocities can be calculated from the Doppler-shifted position of the distribution function relative to the fundamental transition frequency of the excited line. To account for Zeeman effects on the velocities in the direction of the background magnetic field, right- and left-hand circularly polarized light are used independently, and the average frequency shift from the two distribution functions is used, as described in detail by Green et al.\cite{Green_Schmitz_Severn_Winters_2019}. Ion and neutral temperatures are derived by fitting a Maxwellian to the VDF. Further details of this diagnostic and technique are described in \cite{Green_2019,Green_Schmitz_Zepp_2020}.

\subsection{Passive Spectroscopy}
\label{sec:spectroscopy}

There are three spectrometers available for measurement on MAP, any of which can be connected by optical fiber to the sample the yellow line-of-sight in \cref{fig:diagOverview}. The most widely used one is a Thorlabs CCS100 grating spectrometer with a range of 350-700 nm, spectral accuracy under 0.5 nm, and a resolving power of 870. Many prominent argon ion and a few neutral argon lines fall into this range. The integration time can be set as low as 10  $\U{\mu s}$, making this spectrometer very useful for general plasma characterization. The second spectrometer is a Thorlabs CCS175, which has a higher wavelength range of 500-1000 nm, covering the prominent neutral argon lines from 750 to 850 nm. It has a spectral accuracy better than 0.6 nm and a resolving power of 1020.\\

In addition to these compact survey units, MAP has a Princeton Instruments SP2500i grating spectrometer, covering the 250-950 nm range with a resolving power of 7100. This spectrometer uses a thermionically cooled PIXIS 256E CCD camera with a quantum efficiency of $\>$20\%. While capable of much higher resolution, this spectrometer is very slow compared to the Thorlabs units, taking about a minute to sweep the entire spectrum. However, the CCD camera can be replaced with a photomultiplier tube, so that a single line can be measured with great speed.\\

\subsection{Control and Data Acquisition Software}
\label{sec:controlInterface}

The RF power supplies, interferometer, spectrometers, diagnostic positioning stages, data acquisition, and parts of the LIF system are controlled through modular Python-based control interfaces. This allows synchronization of the different MAP subsystems and automation of measurement campaigns. For example, we can move the diagnostics in \cref{fig:diagOverview} into position, trigger a plasma discharge, read out interferometry and spectroscopy from that pulse, and repeat the procedure at the next location. This setup allows for a high degree of automation during MAP's operation and scripting of measurement campaigns with hundreds or thousands of shots. We are currently completing automation of the LIF system.

\section{Overview of Experimental Results}
\label{sec:firstResultsMAP}

In this section, we present an overview of MAPS's experimental results. Some of these findings have previously been published or are currently under review for publication elsewhere. In these cases, we will limit our description to a summary of the main results and provide a reference to these detailed studies.

\subsection{Optical Emissions in Different Modes of Operation}

\begin{figure}[htp]
    \includegraphics[width=\linewidth]{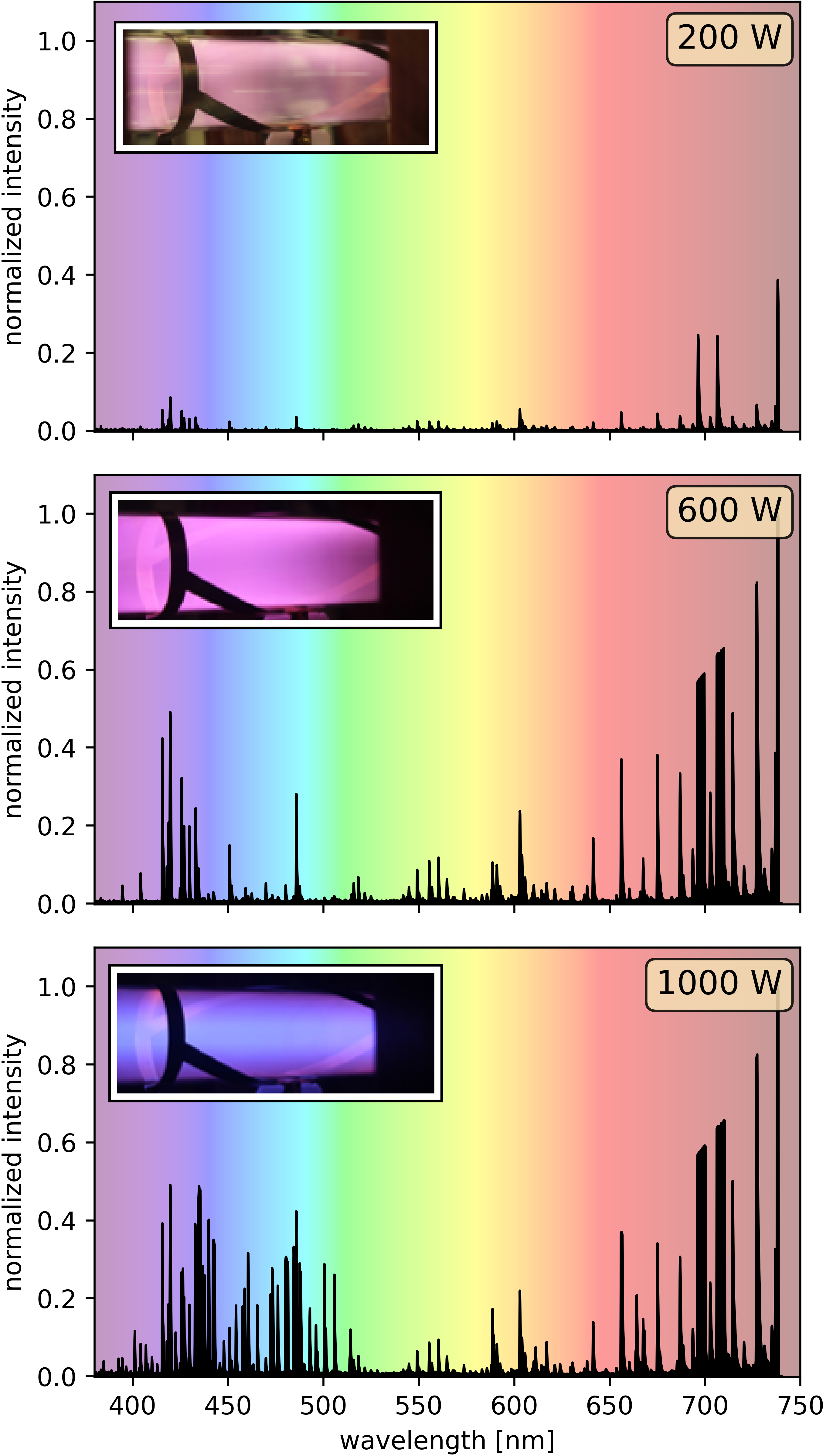}
    \caption{Photos of the antenna region and passive spectroscopy thereof plotted over the visible color range. Shown is the transition from capacitively-coupled (top) through inductively-coupled (middle) to helicon (bottom) plasma in a 1 Pa, 41 mT plasma with a 20 cm long right-helical antenna. Intensities are normalized to the strongest line of the helicon plasma case at 738 nm.}
    \label{fig:specComp}
\end{figure}

Starting from a cold gas with minimal ionization, the plasma transitions from capacitively-coupled, through inductively-coupled to helicon mode as input power increases. This progression is demonstrated in \cref{fig:specComp} through spectroscopy and photos of the plasma in the antenna region. These discharges are excited inside a rightward directed 41 mT magnetic field at an argon fill pressure of 1 Pa using a single 20 cm long right-handed half-helical antenna. The spectra for the three shown power levels were normalized to the highest signal across all wavelengths and power levels, the 738 nm peak at 1000 W. Normalized emission intensities are therefore in the same unit across all three panels.\\

At 200 W RF power (top in \cref{fig:specComp}), plasma emissions are predominantly violet in the 380 to 450 nm region and red past 700 nm, indicating neutral argon excitation. The plasma shows a hollow core when viewed through the axial end ports. This plasma structure and neutral-heavy emission spectrum indicate a capacitively coupled discharge.\\

At 600 W the plasma has transitioned from capacitively- to inductively-coupled (center in \cref{fig:specComp}). Spectroscopy shows an increase of intensity in the violet and red regions of the spectrum, as well as two peaks in the blue region between 450 and 500 nm. However, this blue light is not indicative of ArII and helicon-typical neutral depletion. A comparison with the NIST database\cite{NISTLineData} identifies these lines as very strong FeI lines at 452.59 nm and 487.13 nm. The reason these lines appear after transitioning into the inductive mode is most likely that the plasma is now making significant contact with the stainless steel end sections of the vacuum vessel for the first time.\\

At 1000 W (bottom in \cref{fig:specComp}) the plasma has transitioned into helicon mode and a blue core extends from the antenna to the left into and past the spectroscopy measurement location. The spectrum shows very strong ArII lines at 458.99 and 472.69 nm, as is characteristic of high-density helicon discharges.\\

\subsection{Impedance Jumps during Mode Transitions}

Another characteristic of mode transitions is a strong jump in plasma density and impedance. This can be seen in \cref{fig:heliconApproachTraces}, which shows set-point, forward, and reflected RF power as the plasma transitions into helicon mode for the same operating conditions as in \cref{fig:specComp}.\\

\pic{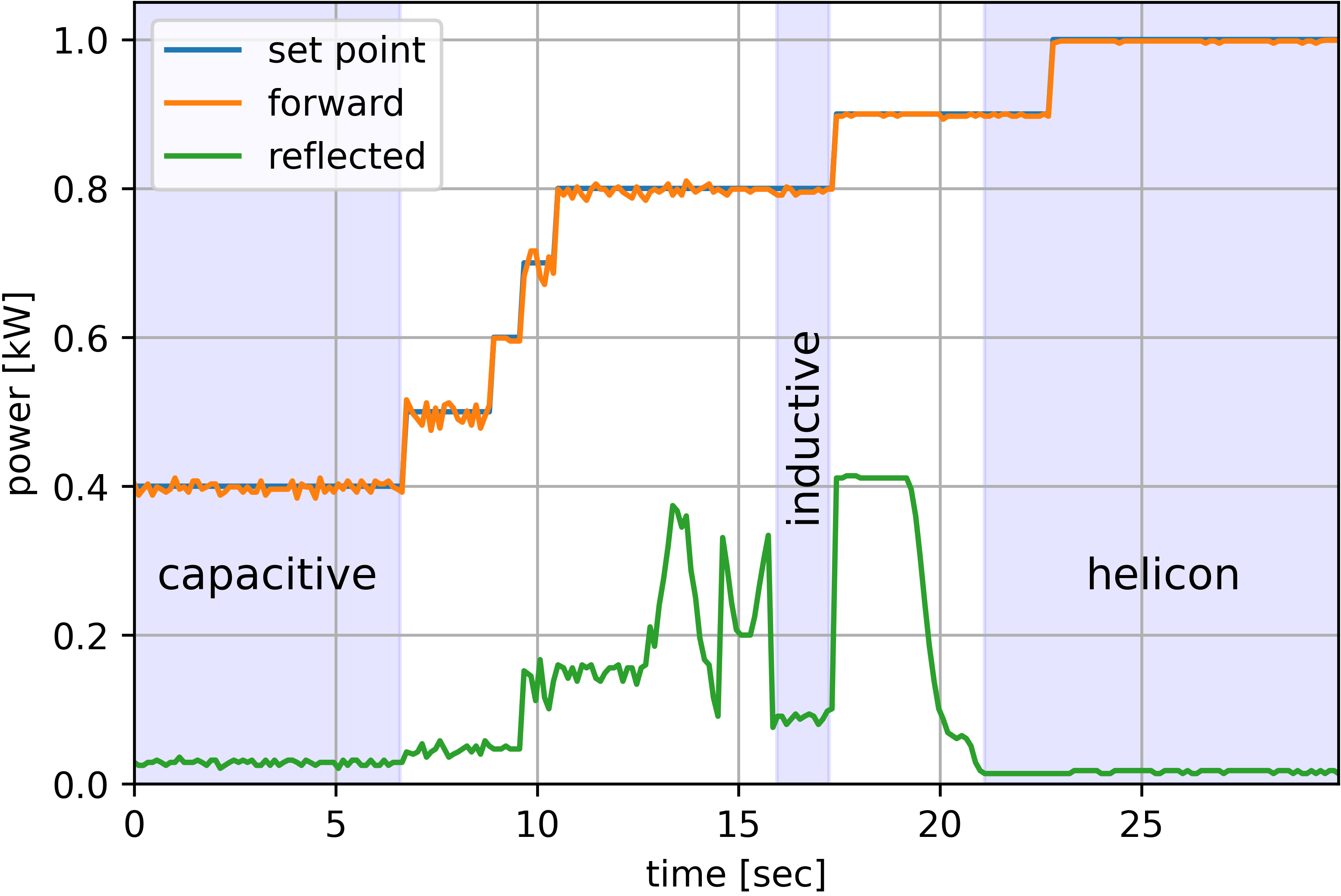}{Set point, forward, and reflected RF power during the approach to helicon plasma. After each mode transition the impedance changes significantly, resulting in an increase of reflected power. By adjusting the impedance matching network we can reduce the amount of reflected power and establish stable operation in each new mode. The impedance is mostly unchanged when increasing RF power while maintaining the same mode, for example when changing from 0.9 to 1.0 kW in the helicon mode.}{fig:heliconApproachTraces}

The discharge starts well-matched at 400 W of total RF input power, sustaining a capacitively coupled plasma. At around 7 seconds the RF power is ramped up to 800 W. Reflected power is initially relatively high at 160 W, indicating an impedance change as the plasma transitions from capacitively to inductively coupled. To compensate, we adjust the matching network, starting at around 12 seconds. At 16 seconds this yields improved matching and reduces reflected power to 100 W. At around 17 seconds the set-point power is increased to 900 W and the reflected power shoots up to more than 400 W as the plasma transitions into helicon mode, accompanied by a massive impedance change. We adjust the matching network again, leading to around 30 W reflected power at 21 seconds. A subsequent increase of the RF power to 1000 W at 23 seconds does not change the impedance further since the plasma stays in helicon mode. These jumps in plasma impedance correspond to the commonly seen jumps in density\cite{chi1999resonant} as the plasma transitions from capacitive, through inductive to helicon mode. The matching network settings required for low-power helicon mode provide good impedance matching up to 10 kW of power with only minor adjustments. This indicates that the helicon plasma impedance changes only weakly as power is increased.\\

\subsection{Positive Density Scaling with Magnetic Field}

The density of a helicon plasma is generally found to increase with the magnetic field. To demonstrate this effect we scanned MAP's field from 25 to 50 mT in 2.5 mT increments. This study was performed at 1.3 kW of RF input power and 1.0 Pa of argon pressure. This scan was performed with a 20 cm long half-helical antenna before changing to a 10 cm long antenna that achieves higher densities as explained later in \cref{sec:antOpt}. Density measurements were taken directly under the antenna. The results in \cref{fig:nBScaling} indicate the expected linear increase with density, specifically we find a slope of $1.6\E{17}\U{\dens\, mT^{-1}}$.

\pic{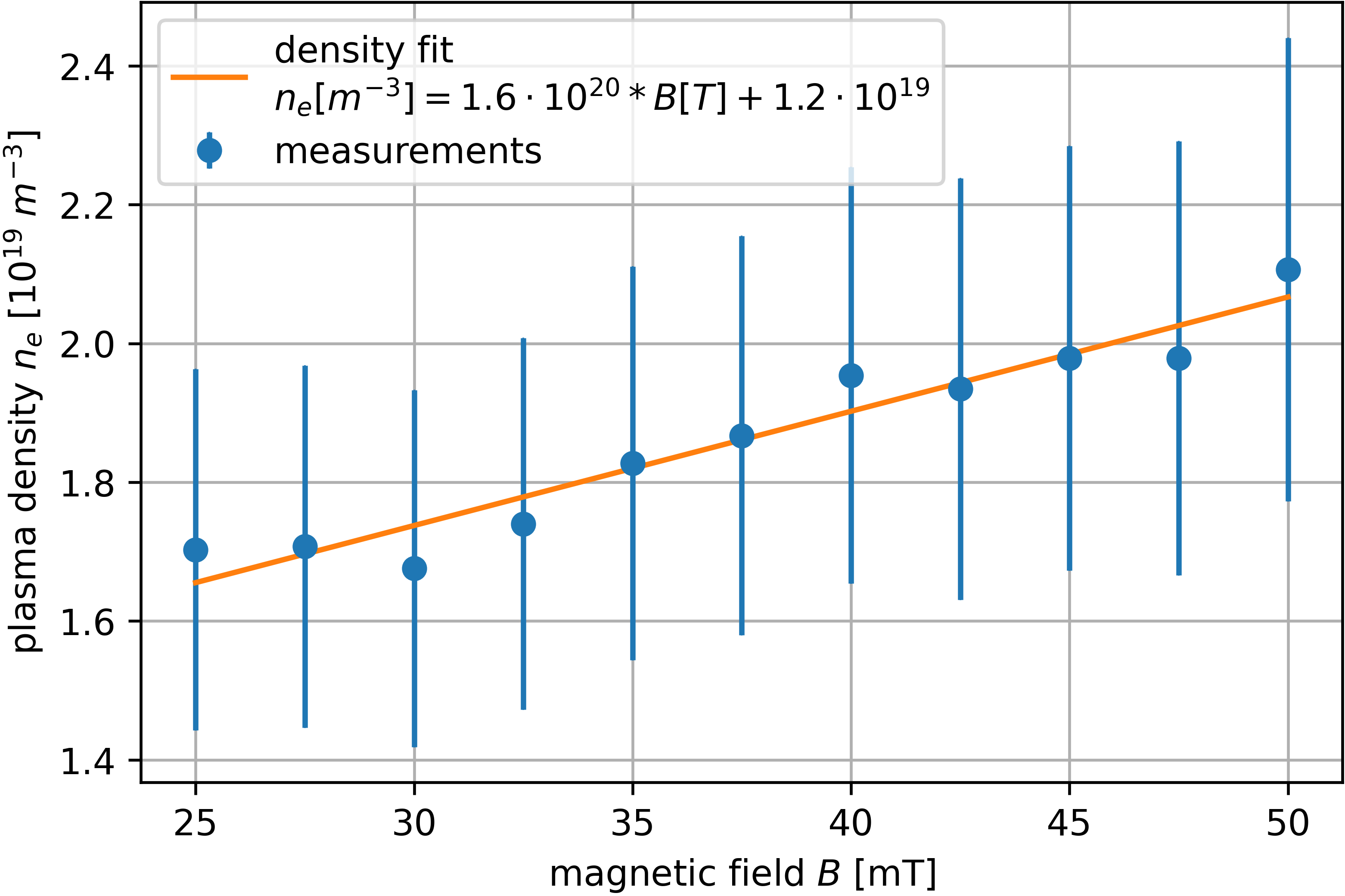}{Density scaling with magnetic field under the antenna in MAP between 25 and 50 mT. The measurements show a linear increase of density with field strength, which is characteristic of helicon plasmas.}{fig:nBScaling}

\subsection{Low-Power 2D Density and Temperature Measurements}
\label{sec:2Dmaps}

Typical low-power MAP operating conditions are 1.3 kW RF power, 1 Pa argon, and 50 mT field, combined with a single 10 cm long, right-handed antenna centered at $z = 0 \U{cm}$ in \cref{fig:comMag}. To characterize this setup, we took measurements at 16 radial positions, spaced 3 mm apart, and repeated this scan at six axial positions. The line of sight was partially obstructed at the antenna location, so measurements there had to stop at a radius of 1.9 mm. 86 measurements yielded an LIF signal strong enough for further processing and interpolation. We averaged LIF scans over multiple shots to bring the uncertainty in the LIF intensity below 1\% in most locations. Areas near the radial boundaries feature a significantly lower ion population, resulting in very low LIF intensity and correspondingly higher uncertainty. In the outermost 0.5 mm, averaging was used to maintain typical uncertainties of 3\%, reaching at most about 10\%.\\

We can derive plasma densities and ion temperatures from these measurements, as shown in \cref{fig:2DDensPlot,fig:2DTempPlot}. In both plots, the antenna location is indicated by the horizontal black lines. The density peaks at $4.5\E{19}\U{\dens}$ about 15 cm downstream of the antenna and has a hollow profile. However, we find that at higher pressures the density peaks on-axis. This mechanism is described in more detail later in \cref{sec:ionSource} and in \cite{Zepp_2024_thesis}.\\

\pic{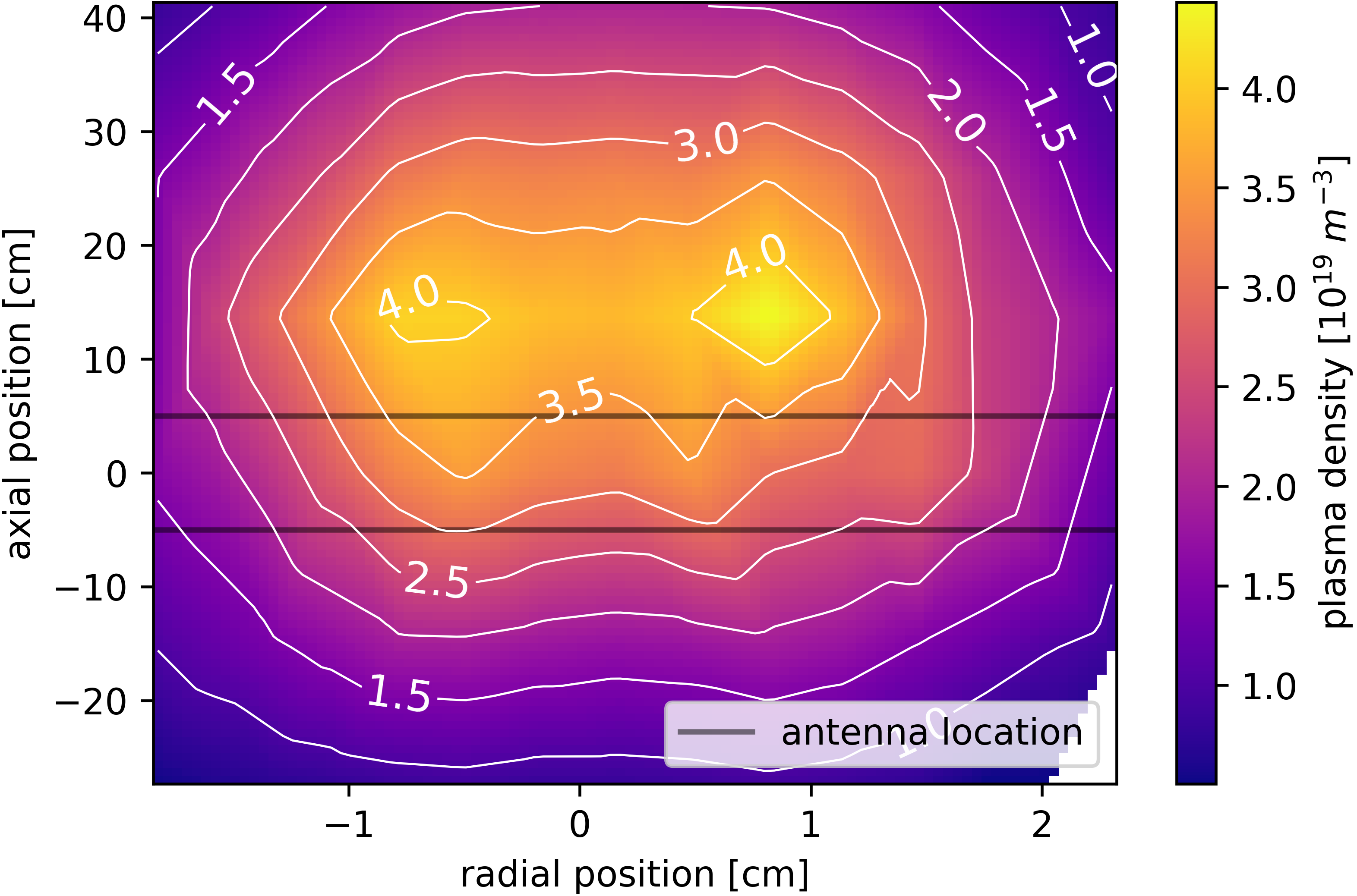}{Plasma density profiles in MAP at 1.3 kW, 50 mT, 1 Pa and with a 10 cm long right-helical antenna. Interpolated from 86 LIF measurements.}{fig:2DDensPlot}

The temperature measurements are significantly noisier than the density measurements. This is because the derivation of temperature is very sensitive to small changes in the peak width of the VDF. We have therefore chosen to cap the color bar in \cref{fig:2DTempPlot} at 1 eV, as higher values all have relative uncertainties greater than 30\%. General trends are nonetheless quite clear. The largest temperatures occur at the radial boundary, reaching 1 eV and likely higher right below the antenna. The ion temperature elsewhere approximately follows the axial density profile and has little radial dependence. Ion temperatures in the bulk plasma reach up to 0.4 eV but are mostly at 0.2 eV or lower.\\

\pic{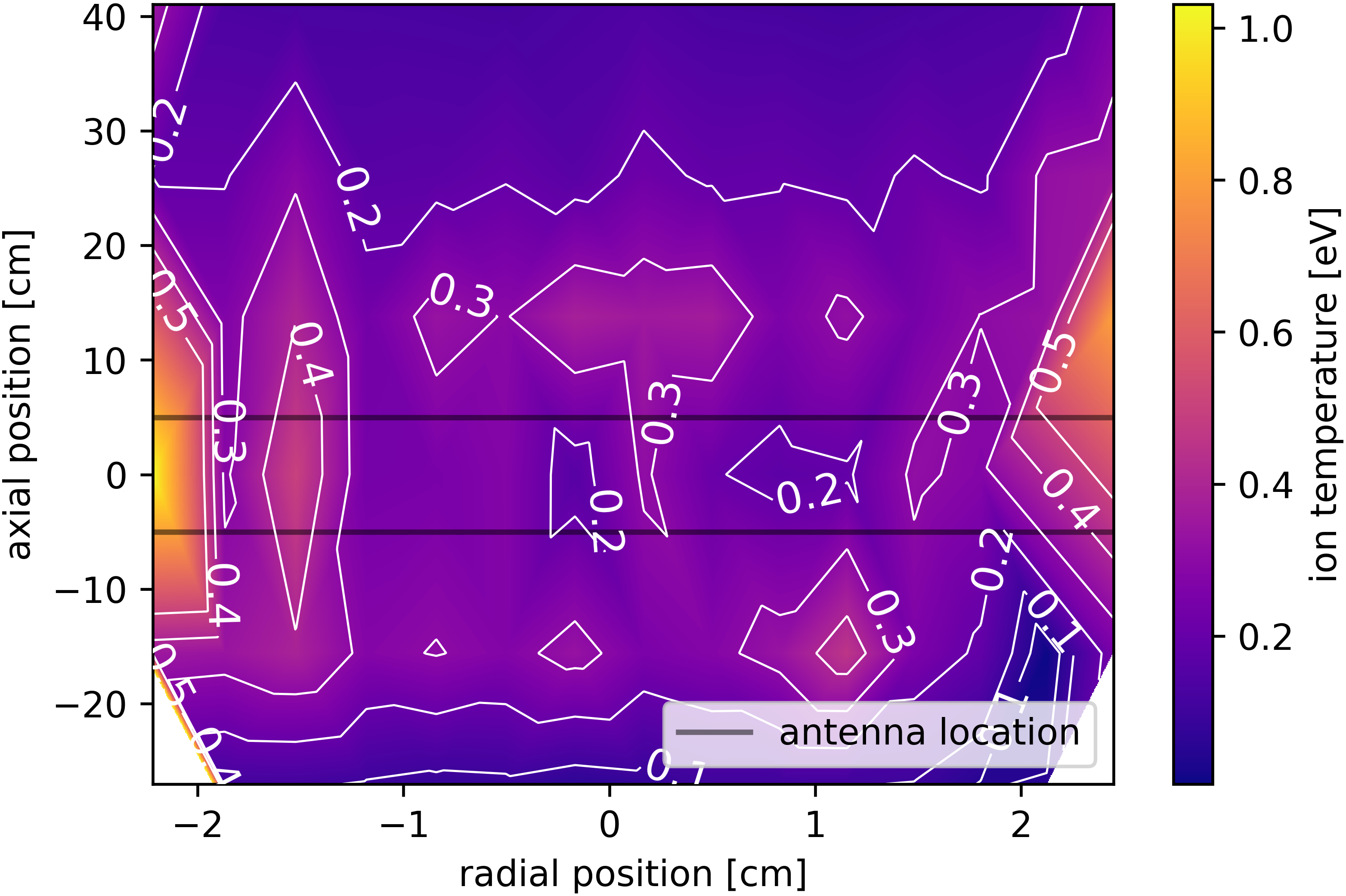}{Ion temperature profiles in MAP for the same conditions as in \cref{fig:2DDensPlot}. Relative errors on temperature measurements are higher than for density measurements. We therefore exclude measurements with a relative error over 30\%.}{fig:2DTempPlot}

\subsection{Increased Plasma Length during High-Power Plasma Operation}

\begin{figure}
    \includegraphics[width=\linewidth]{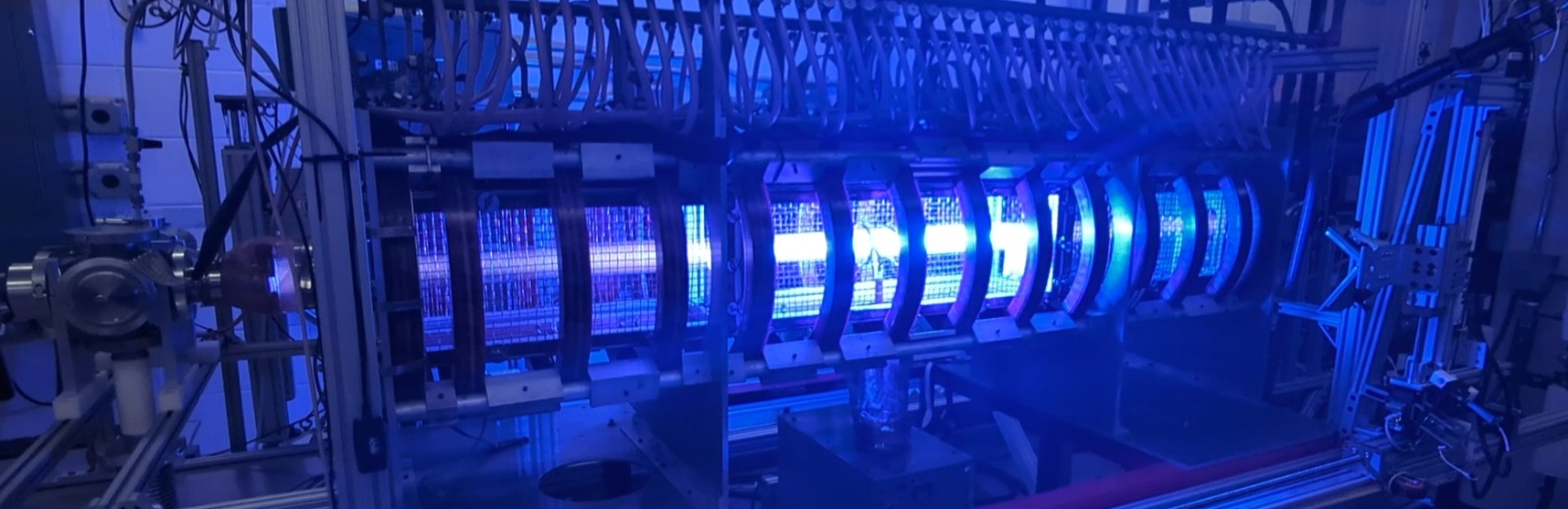}
    \caption{\label{fig:MAPoverview}
    The MAP experiment during a full power, 10 kW pulsed helicon discharge from a single 10 cm right-helical antenna at 3 Pa and 50 mT. The plasma is highly directional, featuring a blue core that extends from the antenna through the entire right half of the vacuum vessel.}
\end{figure}

\Cref{fig:MAPoverview} shows a photo of the experiment during a 10 kW RF power discharge at a pressure of 3 Pa with a 50 mT field. We find that at high power the plasma is much brighter and the blue core extends throughout the length of the magnetized part of the vacuum vessel in the direction of preference as described later in \cref{sec:plasDir}. In general, the plasma core extends further as power levels are increased.\\

Currently, quantitative measurements at these higher powers are limited. We find that the $3$d$^4$F$_{7/2}$ ArII state used for LIF is strongly depleted at these high power levels, which makes LIF unavailable with the current setup. At the same time, on-axis plasma densities even at medium power levels reach the $\ebd{20}$ range, as shown later in \cref{sec:powerScaling}. At these densities, the interferometer beam experiences significant refraction and eventually cuts off as described in \cref{sec:interferometer}. We plan on enabling measurements in this high-power regime by upgrading the interferometer source from 105 GHz to a higher frequency. 

\subsection{Multi-Antenna Operation at Increasing Power Levels in the $\ebd{20}$ Density Range}
\label{sec:powerScaling}

\begin{figure*}[htp]
\centering
\includegraphics[width=0.9\textwidth]{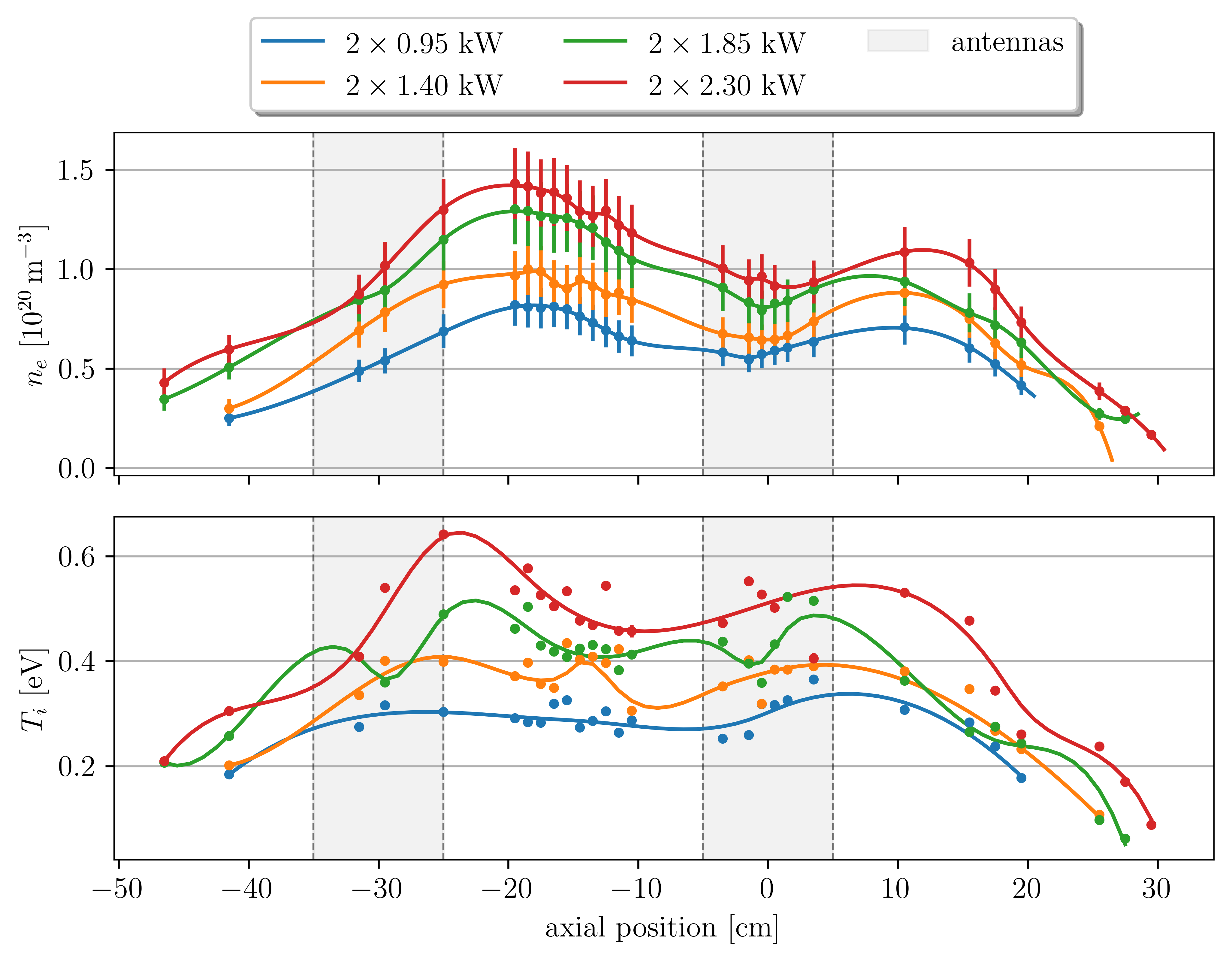}
\caption{Core plasma density $n_e$ and ion temperature $T_i$ in a dual antenna setup at increasing power levels from 0.95 to 2.3 kW per antenna. The 10 cm long antennas are the axial locations indicated by grey shading. Top: The density increases with increasing power levels. The inter-antenna region is almost completely in the $\ebd{20}$ region at $2\times2.3\U{kW}$. Bottom: Ion temperatures reach up to 0.65 eV and peak at the right edge of each antenna.} 
\label{fig:powerScan}	
\end{figure*}

One of the most significant areas of MAP research with regard to the AWAKE project is the absolute density and homogeneity of the plasma core in a multi-antenna configuration. We investigated such a setup with two 10 cm long right-handed antennas, spaced 30 cm apart in the central 50 mT region on MAP. We set the fill pressure to 3 Pa and the relative phase difference at the RF generators to zero, with total power levels ranging from 1.9 to 4.6 kW, distributed evenly across both antennas. The results for plasma density and ion temperature are shown in \cref{fig:powerScan}.\\

Densities between the antennas are in the high $\ebd{19}$ region even at 1.9 kW and reach up to $1.4\ed{20}$ at 4.6 kW. Importantly, these power levels represent only 23\% of the total available power. Focusing on the region between antenna centers, namely between -30 cm and 0 cm in \cref{fig:powerScan}, allows us to calculate an average density and standard deviation in this region at different RF power levels. This analysis is shown in \cref{fig:PowerScaling}.\\

\pic{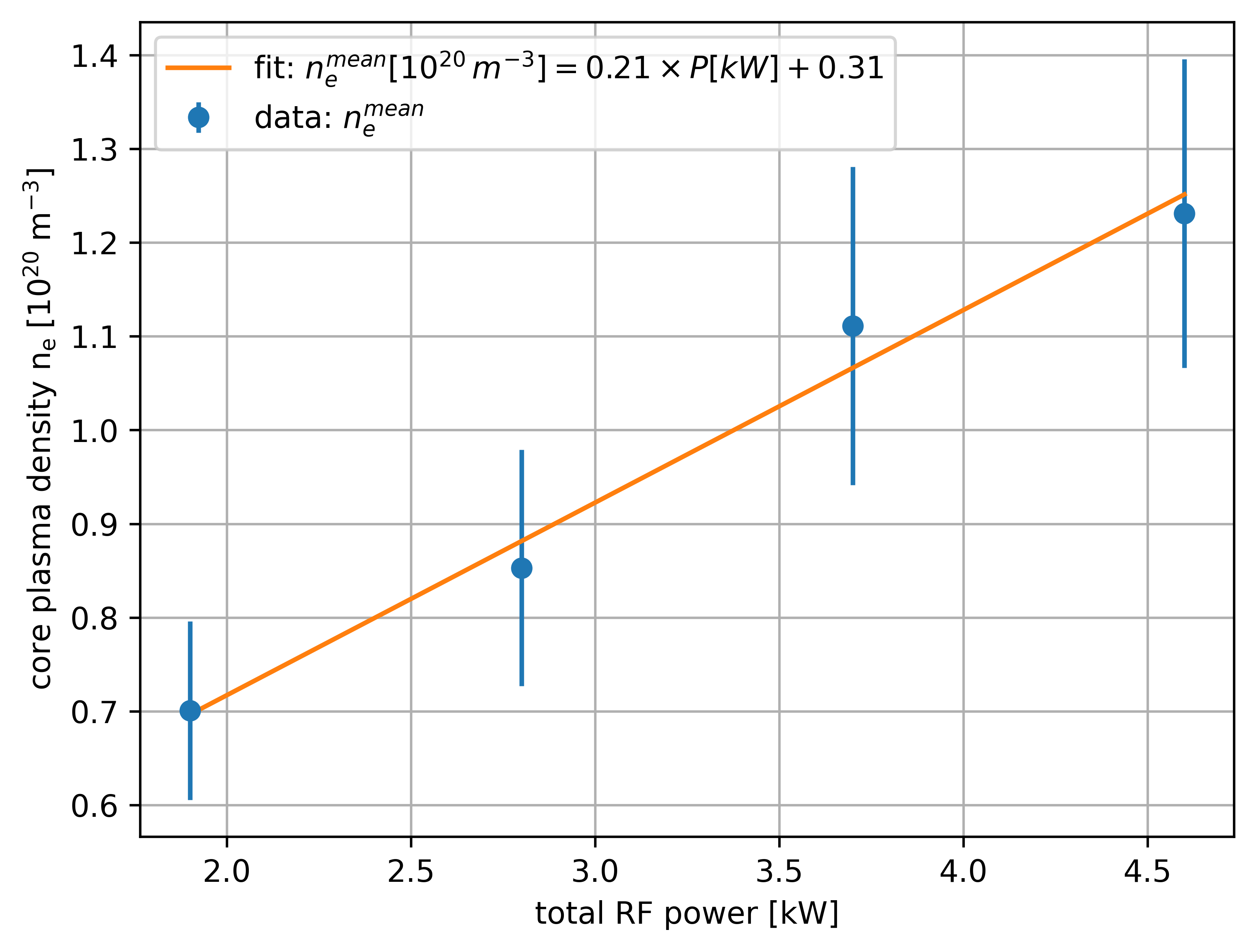}{Mean core plasma density between the antenna centers in \cref{fig:powerScan} in dependence on total power levels. The density increases linearly with total RF power, reaching $1.2\ed{20}$ at 4.6 kW.}{fig:PowerScaling}

The average density in the inter-antenna region is $7.0\ed{19}$ at 1.9 kW and reaches $1.2\ed{20}$ at 4.6 kW. Axial density variations range up to 30\%. We find the following linear relationship between mean density $ n_e^{mean}$ and RF power $P$ in this range:

\begin{align}
\label{eq:densScaling}
    n_e^{mean}[\U[]{\ebd{20}}] &= 0.21 \times P[\U[]{kW}] + 0.31.
\end{align}

However, as the RF input power increases the ion temperature rises significantly as shown on the bottom panel in \cref{fig:powerScan}. Temperatures in the inter-antenna region region range from 0.4 eV at 1.9 kW to 1.3 eV at 4.6 kW. These higher ion temperatures are likely to coincide with raised electron temperatures. The latter would lead to increased line radiation which in turn would flatten the density scaling predicted by \cref{eq:densScaling}.


\subsection{The Mechanism Determining Plasma Directionality and Preference of Right-Handed Helicon Modes}
\label{sec:plasDir}

\begin{figure*}[htp]
\centering
\includegraphics[width=1.0\textwidth]{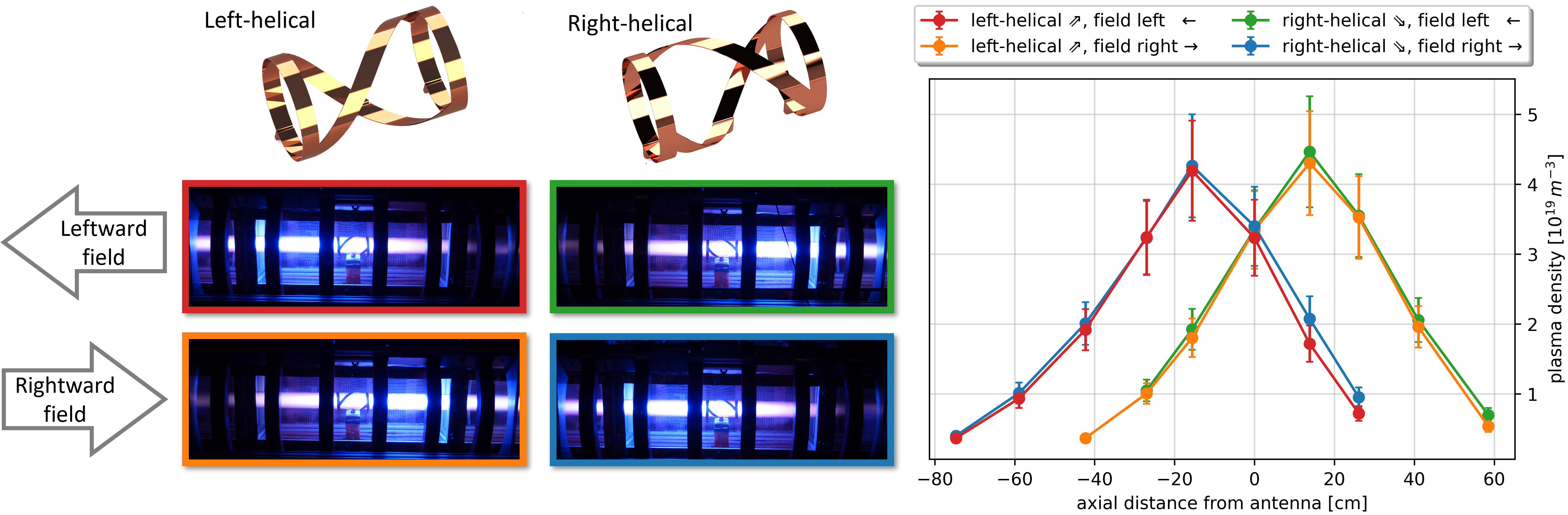}
\caption{Directionality of the MAP plasma in dependence on the antenna helicity and magnetic field direction. The left and right columns show discharges with a left- and right-helical antenna, respectively. Axial magnetic fields in the top row point to the left and to the right in the bottom row. Reversing either the antenna helicity or field direction results in a reversal of the discharge. The density measurements on the right show an exact mirroring of the discharges, in agreement with the photographs on the left. This effect is tied to the preference for right-handed helicon modes and arises due to the interaction of the background field with the radial plasma density gradients as explained in \cite{granetzny2023}.} 
\label{fig:directional}	
\end{figure*}

Using the flexible antenna setup, modular Faraday screen, and field reversal capability described in \cref{sec:antennas,ssec:faraday,sec:magSetup}, we were able to study helicon discharges for leftward and rightward magnetic fields in combination with left- and right-helical antennas on MAP. \Cref{fig:directional} shows four different MAP discharges at 1.3 kW RF power, 1 Pa argon fill pressure and 50 mT magnetic field using a 10 cm long half-helical antenna. In all cases, the plasma contains a very bright blue core, which is imaged blue-white on the CCD due to detector saturation. Importantly, this core is not symmetric around the antenna but shows a clear directionality, extending much further to the left or right depending on the combination of background field direction and antenna helicity.\\

The discharge on the top left in \cref{fig:directional} (red frame) is created using a left-handed antenna inside a magnetic field pointing to the left. In this configuration, the discharge is directed leftward. On the bottom left (orange frame) all parameters are identical, except that the magnetic field direction has been reversed. This reversal leads to a rightward discharge. The configurations on the right (green and blue frames) feature the same magnetic fields as on the left but use a right-handed antenna. This setup leads to reversals of the discharge direction compared to the setups with a left-handed antenna. Density measurements, shown on the right in \cref{fig:directional}, further show that the density profiles follow the optical emissions, are highly asymmetric, and are exactly mirrored around the antenna location in all four cases. Overall, we find that the discharges are strongly directional and that the discharge direction can be flipped by reversing the field or antenna helicity.\\

Starting from these observations, we were able to show computationally and analytically that this directionality is tied to the long-known preference of right-handed helicon modes\cite{light1995helicon} and that both effects arise from the interaction of the radial plasma density gradient with wavefield currents and background field. The full results of this study were published previously by Granetzny et. al. in \cite{granetzny2023,granetzny2024}. Absolute density values in \cref{fig:directional} are higher than in this earlier study since the new microwave interferometer allowed us to recalibrate the scaling law as described in \cref{sec:interferometer,sec:LIF}. This full study makes extensive use of a new COMSOL-based finite-element model that can calculate the helicon wavefield and power deposition patterns in MAP. This model is described in detail in \cite{granetzny2025comsol}.\\

Lastly, such forced discharge reversals represent a new method to establish that a device operates in helicon mode. Since helicon waves are magnetized plasma waves, it is physically intuitive that they respond strongly to a reversal of the magnetic field direction. In particular, a capacitively- or inductively-coupled plasma would not be influenced by a reversal of the background field. To our knowledge, this fact has not been emphasized in the literature which typically uses density jumps to establish that a plasma source operates in helicon mode. A plasma direction reversal induced by a magnetic field or antenna helicity flip can therefore serve to demonstrate helicon mode operation and has the added benefit of not requiring any density diagnostics. However, the magnetic field has to be sufficiently similar on both sides of the antenna.\\


\subsection{2D Ionization Source Rate and Implications for Plasma Homogeneity and Density}
\label{sec:ionSource}

\begin{figure*}[htp]
\centering
\includegraphics[width=0.9\textwidth]{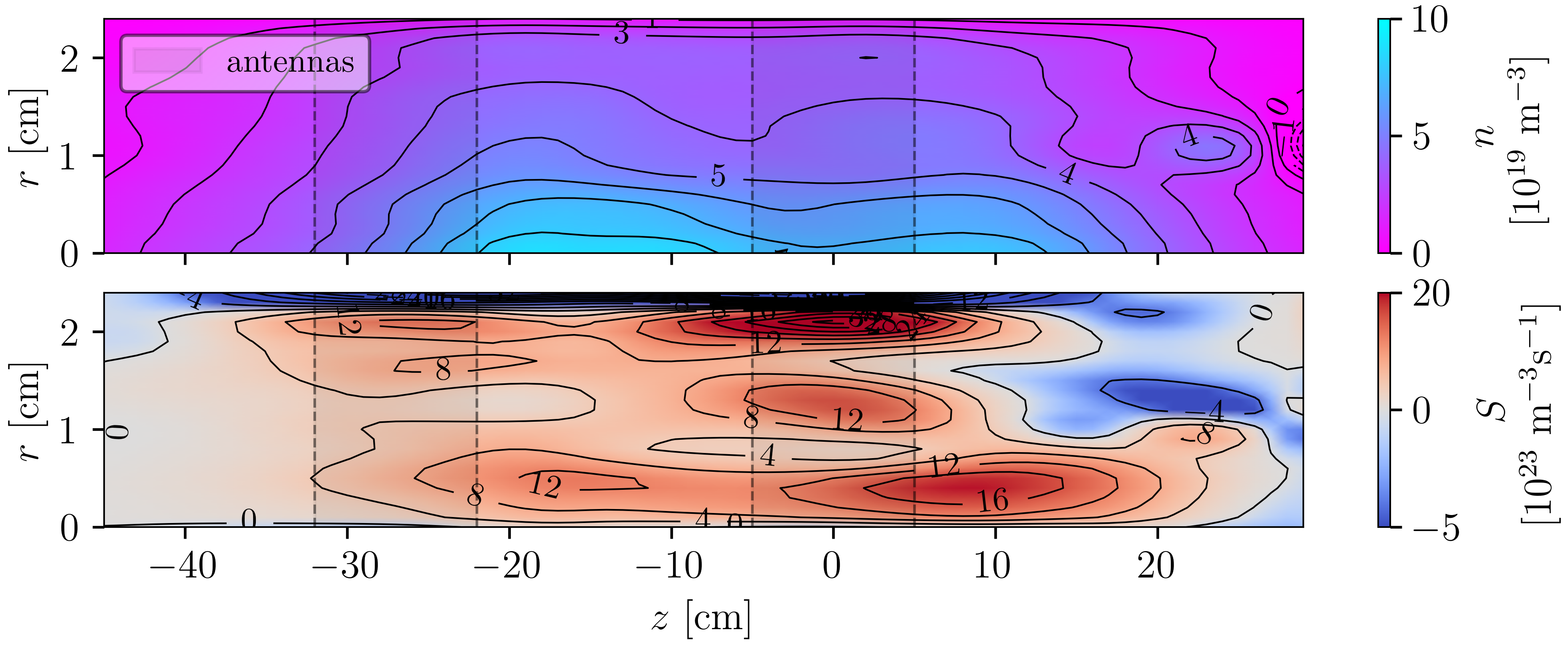}
\caption{Density profiles and ionization source rates in dual antenna operation. Top: Density profiles in a setup with two 10 cm long right-helical antennas at 1 kW each at 3 Pa in a 50 mT field pointing to the left. The plasma shows a relatively smooth on-axis density around $8\ed{19}$. Bottom: Particle source rate calculated density and flow velocity measurements. There is an ionization front at $z=10 \U{cm}$ as the plasma encounters the incoming neutral gas. However, the source rate elsewhere is dominated by sources close to the wall where ions are recycled.  } 
\label{fig:dualAntenna}	
\end{figure*}

We used MAP to study the particle source rate profiles at various powers and pressures. This was possible by measuring the axial and radial ion flow velocities along with the plasma density using LIF and interferometry. Under the assumption of axisymmetry, these measurements allow for direct calculation of the source rate from the continuity equations. \Cref{fig:dualAntenna} shows an example of such a measurement for two 10 cm long right-helical antennas, each at 1 kW, in 3 Pa of argon.\\

We find that in the plasma core, the source rate is high in regions of high plasma density. The source rate peaks near the density peak. However, the source rate is also high in the low-density region directly under each antenna. These additional source rate peaks suggest the presence and strong damping of Trivelpiece-Gould waves propagating radially inwards from the antenna.\\

The source rate measurements also showed that the contribution to the source rate calculated from radial flux is much higher than the contribution calculated from axial flux. The radial flux divergence is large in MAP because of the relatively small chamber diameter, as ionized particles reach the wall quickly, recombine, and are recycled into the plasma as neutrals to maintain the high source rate and radial flux. \\

The results from this work suggest that small chamber diameters can be utilized to alter the fueling mechanism of helicons. A small diameter chamber, such as MAP, increases axial homogeneity since there is far less axial dependence than radial dependence. In addition, a close chamber wall allows for much easier sourcing of neutrals for ionization in the plasma core, resulting in higher density. This work has been published by Zepp et. al. in \cite{Zepp_2024}.\\


\subsection{Impact of Plasma and Neutral Flow Direction on Density}
\label{sec:flowDir}

As discussed earlier in \cref{sec:plasDir}, we can choose antenna helicity and magnetic field direction such that the plasma flows either to the left or right in \cref{fig:MAPCAD}. In addition, by opening the gas line inlet valves on the left or right, we can inject argon directly above the vacuum pump or force it through the entire vacuum vessel. Using these capabilities allows us to study how the plasma and neutral flow directions impact the plasma density. It is important to note, that all measurements presented so far used a forced gas flow.\\

\begin{figure*}[htp]
\centering
\includegraphics[width=0.9\textwidth]{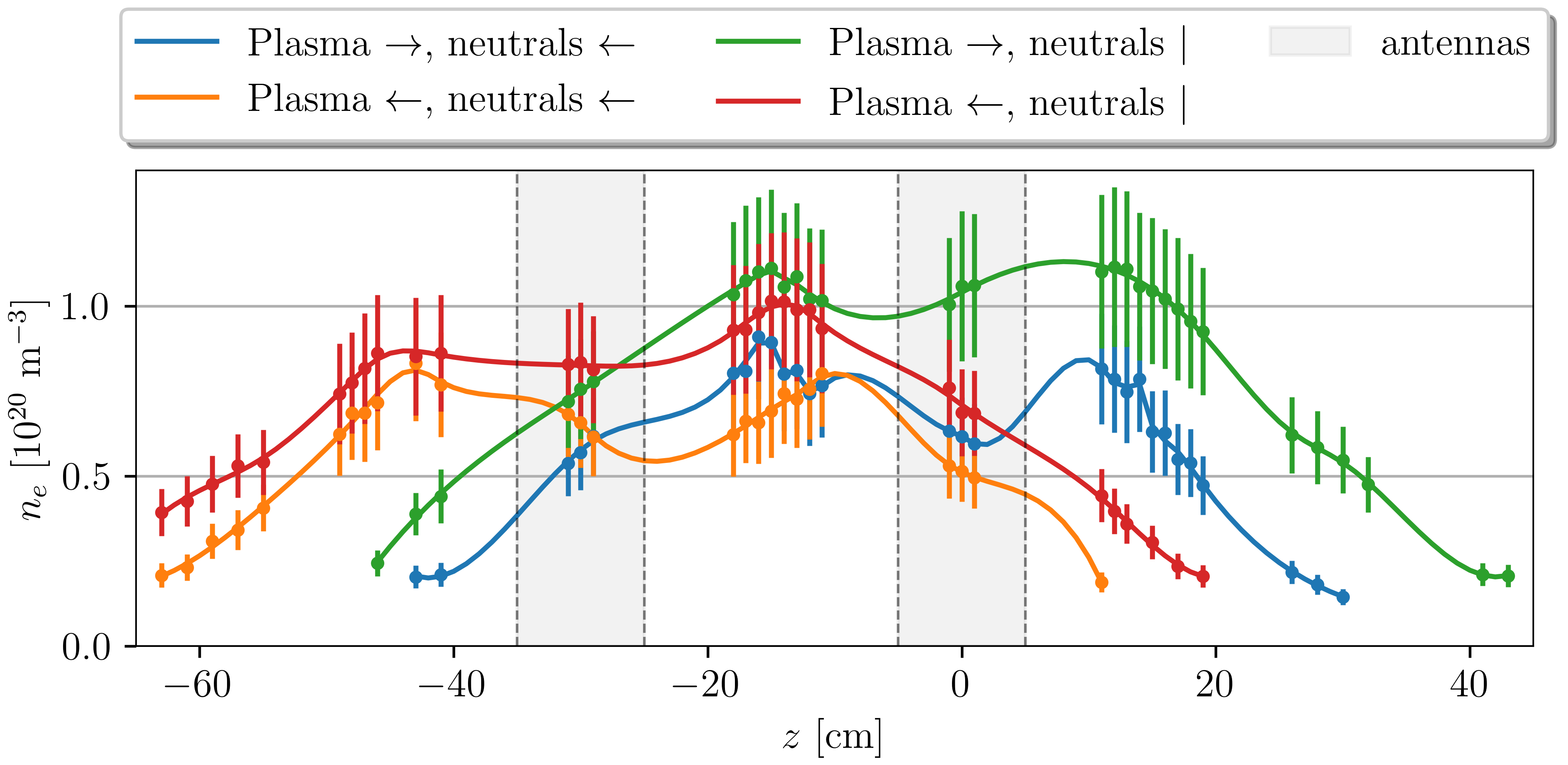}
\caption{Impact of argon ion and neutral flow directions on the on-axis density profiles. When neutral gas is forced through the vacuum vessel (blue and orange cases) a plasma flow reversal leads to a profile mirroring about the center between the antennas at $z = -15\U{cm}$. This result resembles the single antenna studies in \cref{sec:plasDir}. Configurations without a forced background neutral flow (red and green cases) exhibit both higher and more uniform density profiles.} 
\label{fig:flowEffect}	
\end{figure*}

\Cref{fig:flowEffect} shows the axial density profiles for four different combinations of neutral and plasma flow. All four configurations use 3 Pa of argon, a 50 mT field, and two 10 cm long half-helical antennas spaced 30 cm apart. In the blue and orange configurations, neutrals are forced to flow from right to left. In the green and red cases, neutral gas enters and is pumped out on the left and merely diffuses into the rest of the vacuum vessel. We will refer to the latter cases as having no neutral flow. In the blue and green cases the plasma flows to the right, whereas in the orange and red cases, it flows to the left.\\

We find that in the cases with background neutral flow, a reversal of the plasma flow direction results in an axial mirroring of the density profiles. This mirroring occurs around the central point between both antennas at -15 cm. These results resemble the single antenna studies in \cref{sec:plasDir}. Importantly, both configurations have comparable absolute densities and axial density variations.\\

In contrast, the configurations without background neutral flow exhibit both higher and more uniform density profiles. In the case of a rightward plasma flow without neutral flow (green case in \cref{fig:flowEffect}) the measured density exceeds $\ebd{20}$ over approximately 35 cm. This is a significant performance improvement over the forced flow configuration studied previously in \cref{sec:powerScaling} and represented here by the blue curve. In particular, we find that the density scaling derived from the forced flow configuration, \cref{eq:densScaling}, predicts a power requirement of 3.3 kW to achieve an average density of $1\ed{20}$, whereas this improved setup requires only 2 kW to achieve the same average density.\\

In addition to the density profiles shown here, we took flow velocity data to calculate the 2D ionization source rate and axial momentum balance for the configurations in \cref{fig:flowEffect}. Our analysis indicates that the configurations without neutral flow exhibit preferable density profiles largely due to the axial momentum transfer between neutrals and ions. The neutrals in the configurations with neutral flow tend to dampen the axial ion flow, leading to more peaked axial profiles. We further conclude, that this effect is present in helicons operating with a single antenna but is much more pronounced in multi-antenna configurations.  This work is being prepared for publication by Zepp et. al. in \cite{zepp2025}.

\subsection{Computational and Analytical Optimization of Power Coupling through Antenna Length Shaping}
\label{sec:antOpt}

\begin{figure}[htp]
\includegraphics[width=\linewidth]{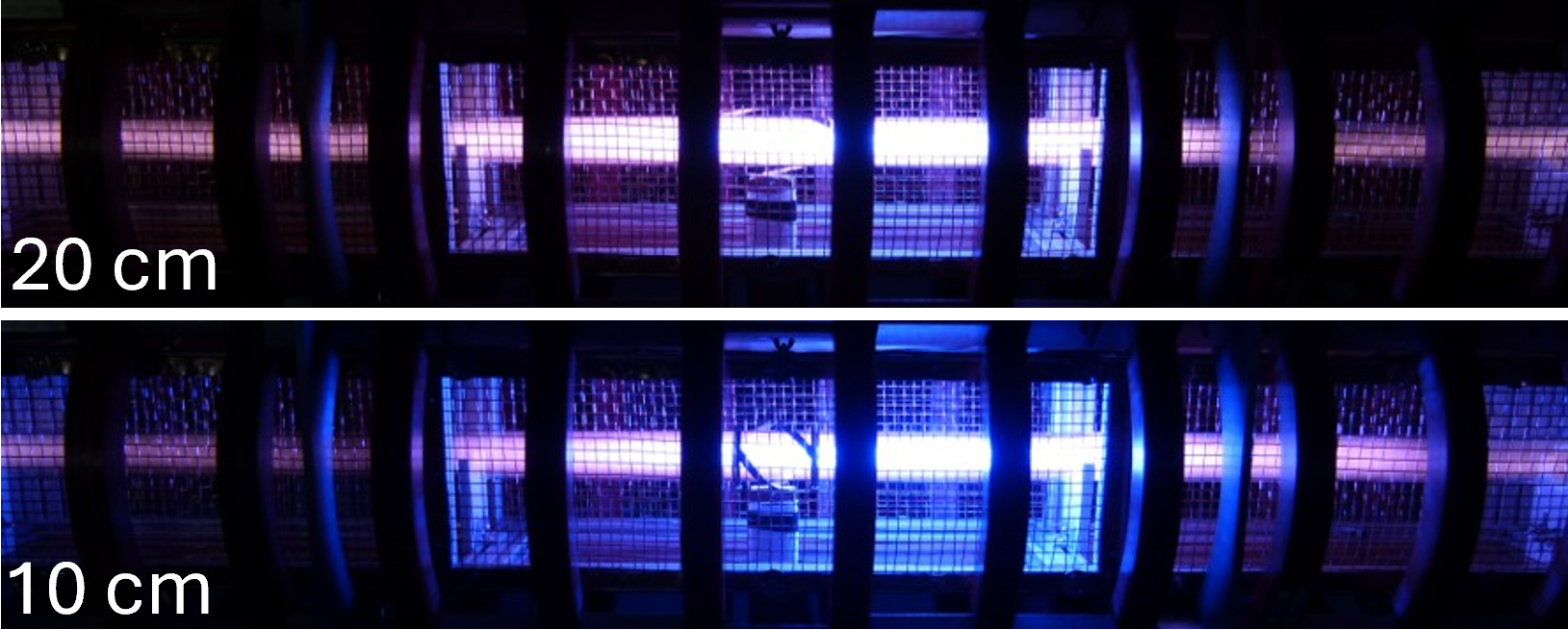}
\caption{\label{fig:lengthExp}Comparison of MAP discharges using antennas of two different lengths with otherwise identical operating conditions at $50 \U{mT}$, 1 kW RF power and an argon fill pressure of 1 Pa. The top panel shows operation with a 20 cm antenna. The discharge is purple and axially symmetric, indicating mainly neutral gas rather than ion excitation. The bottom panel shows operation with a 10 cm antenna. The plasma has a strongly directional blue core, indicating helicon operation and good power coupling.}
\end{figure}

While operating MAP we found that the plasma density and extent of axial density asymmetry are highly dependent on the length of the antenna. An example of this effect is shown in \cref{fig:lengthExp}. Both images show a plasma at 1 kW RF power at 1 Pa of argon in a leftward-directed 50 mT field. The top panel shows operation with a 20 cm long antenna. The discharge is only symmetric and emits mostly purple light, indicating mainly excitation of neutrals as mentioned earlier in \cref{sec:firstResultsMAP}. In contrast, the bottom panel shows the discharge with a 10 cm antenna. Emissions are mostly blue and attributable to argon ions. The plasma is highly directional, favoring the right side as expected from our analysis in \cref{sec:plasDir}.\\

This strong influence of the antenna length can be explained by comparing the antenna power spectrum with the helicon dispersion relation. A helical antenna will excite an axial wavenumber spectrum whose dominant peak depends on the length of the helical part of the antenna, which is the part of the antenna between the two end hoops in \cref{fig:flexAntenna}. At the same time, the helicon dispersion relation\cite{Chen2015} allows only a certain range of wavenumbers at any given RF frequency, plasma density, and magnetic field strength. Antennas of different lengths will have a power spectrum with full, partial, or no overlap with this permissible wavenumber range.\\

\begin{figure}
\includegraphics[width=\linewidth]{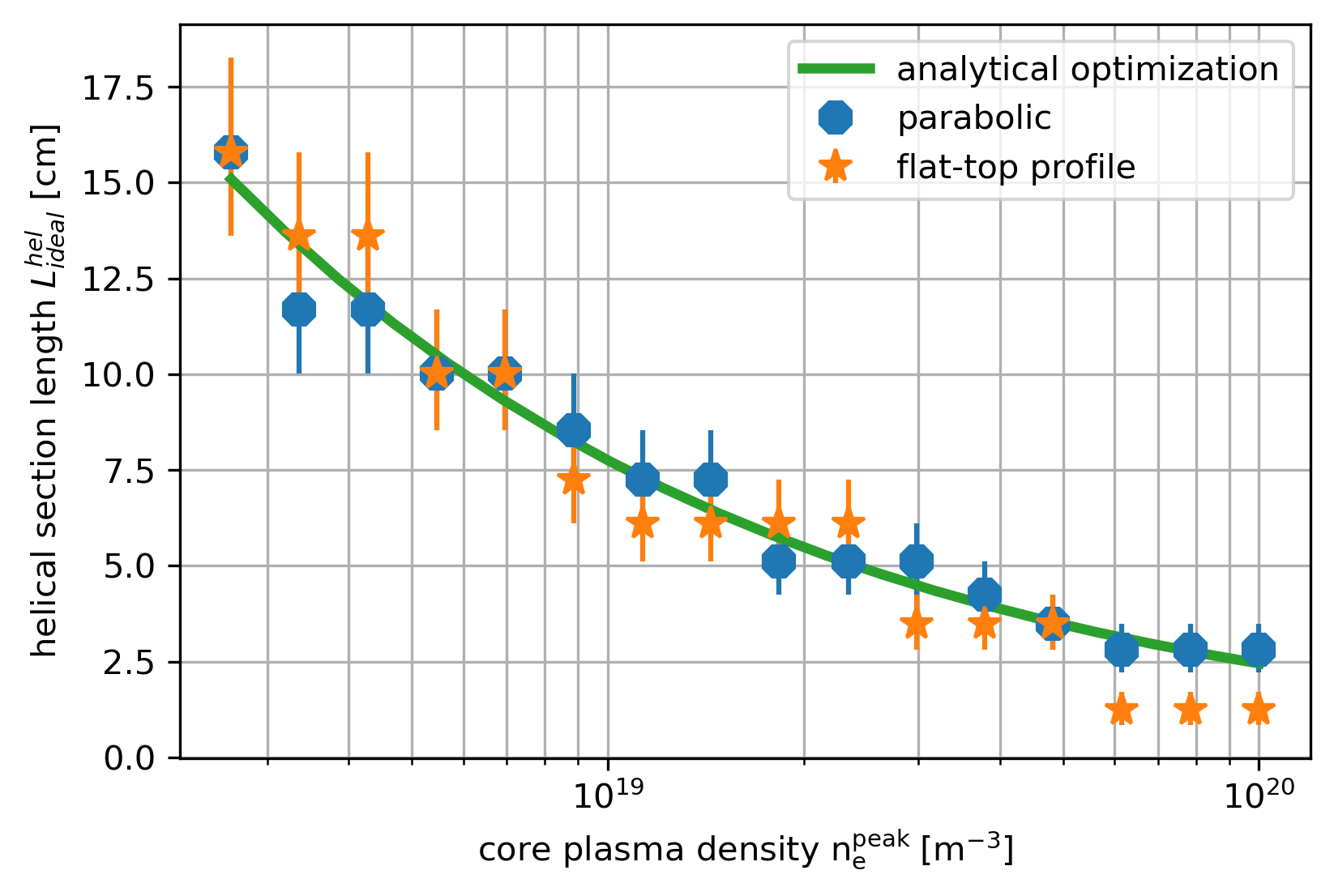}
\caption{\label{fig:ratioComp}Summary of simulation data showing the ideal length of the helical part of the antenna for radial density profiles with parabolic and flat-top shapes. The optimal length is the same or nearly the same in both configurations and can be predicted from \cref{eq:Lideal} by engineering a power spectrum peak that has maximal overlap with the dispersion relation at a given density as shown by the green curve.} 
\end{figure}

We investigated this issue numerically, using a new 3D COMSOL model that can calculate the helicon wavefields and power deposition in MAP. As explained in \cref{sec:plasDir} and \cite{granetzny2023}, helicon plasmas driven by helical antennas are inherently asymmetrical due to the interaction of the wavefields with the radial density gradient. The degree of axial power deposition asymmetry can therefore serve as a proxy for the efficiency of power coupling into the plasma. We analyzed this asymmetry computationally for 800 combinations of antenna length and core plasma density. This study was performed for radial density profiles with both parabolic and flat-top shapes. This enabled us to map out the ideal antenna length for core densities between $\ebd{18}$ and $\ebd{20}$. The main results of this analysis are shown in \cref{fig:ratioComp}.\\ 

The ideal antenna length is nearly identical for both radial density profiles. The green curve shows a reproduction of this relationship based on an analytical comparison between the antenna power spectrum and the helicon dispersion relation. This analysis yields the ideal length of the helical antenna section $L^{helical}_{ideal}$ for a plasma with core density $n_e$ as

\begin{align}
\label{eq:Lideal}
    L^{helical}_{ideal} &= \frac{\pi}{k_w\br{\sqrt{0.61\delta} + \frac{0.61}{\sqrt{1-\delta}}}}\\
    k_w &= \sqrt{\frac{2\pi f n_e \mu_0 e}{B}},\quad \delta = \frac{2\pi f m_e}{e B}.
\end{align}

The full study containing a detailed description of the COMSOL model and analytical derivation has been submitted for publication and can be accessed at \cite{granetzny2025comsol}.


\section{Summary}
\label{sec:summary}

The \textit{Madison AWAKE Prototype} (MAP) was built to develop a high-density, high-uniformity helicon plasma source for use in a beam-driven plasma wakefield accelerator. MAP features a unique combination of magnetic field uniformity, high-power operation with multiple antennas, and overall flexibility and automation. MAP's main diagnostics are heterodyne interferometry, laser-induced fluorescence, and passive spectroscopy, all mounted on computer-controlled multi-axis positioning platforms. This setup allows detailed measurements of plasma density, ion and neutral flows, and temperatures. These experimental capabilities are complemented by a 3D COMSOL finite-element model that can calculate wavefield and power deposition patterns. Using these capabilities we discovered multiple techniques to shape, direct, and homogenize a helicon plasma while increasing its density. At the same time, these studies have resolved several questions in fundamental and applied helicon physics.\\

We were able to reproduce the well-known axial asymmetry of helicon discharges and found that this plasma directionality can be reversed by changing the RF antenna helicity or magnetic field direction. This discovery allowed us to explain for the first time the preference of right-handed helicon modes, which we found to arise from the interaction of the radial density gradient with the helicon wavefields.\cite{granetzny2023} In addition, the degree of plasma directionality can serve as a proxy for efficient RF power coupling. Starting from this observation, we analyzed the impact of antenna length on the power coupling efficiency both computationally and analytically over two orders of magnitude in density. We found that the antenna length can be engineered to optimize power coupling for a given target plasma density and that the ideal length follows a simple analytical expression.\cite{granetzny2025comsol}\\

High-resolution measurements of plasma and neutral densities and flows enabled us to reconstruct the 2D ionization source distribution in MAP. These measurements reveal a significant ionization front where the plasma ends and collides with the neutral flow. However, the source rate in the bulk of the plasma is dominated by recycling at the radial wall. This observation points towards the possibility of smoothing the axial density profile by employing small-diameter vacuum vessels.\cite{Zepp_2024} In addition, we investigated the impact of plasma and neutral flow directions on the axial density profiles. We found that both co- and counterstreaming flows lead to comparable plasma density values on the axis. However, the density increases by half when injecting and pumping gas on the same side of the experiment while directing the plasma flow towards the other side.\cite{zepp2025}.\\

By measuring the on-axis density at increasing power levels, we were able to derive a simple density scaling law that predicts a density of $4\ed{20}$ at 20 kW of total RF power in a dual antenna setup. Such a density would result in an accelerator gain of  $2\U{GeV\,m^{-1}}$, twice the target for the AWAKE project. However, the current variation in axial density is much higher than tolerable in a wakefield accelerator.\\

Optimizing both density and homogeneity at full RF power will be the focus of future research on MAP. The first step in this direction will be to test a coupling-optimized antenna together with the favored combination of plasma and neutral flow direction. We plan to integrate a third RF generator and upgrade the interferometer with a terahertz source to shift the cut-off density into the mid $\ebd{21}$ region. These upgrades will allow us to investigate the low to mid $\ebd{20}$ density space in a variety of triple-antenna setups. Of particular interest will be the impact of different antenna lengths, spacings, and phasings, along with different fill pressures.

\section*{Acknowledgements}
The research presented here was funded by the National Science Foundation under grants PHY-1903316 and PHY-2308846 as well as the College of Engineering at UW-Madison.

\section*{References}
\bibliography{references}

\end{document}